\documentclass[apj]{emulateapj_rtx4}

\usepackage{mathptmx}
\usepackage{epsfig}
\usepackage{natbib}
\slugcomment{\rm To Appear in the {\sl Astrophysical Journal}}
\shortauthors{Eracleous, Hwang, \& Flohic}
\shorttitle{Energy Budgets of LINERs}

\def\aj{\rm{AJ}}                   
\def\araa{\rm{ARA\&A}}             
\def\apj{\rm {ApJ}}                
\def\apjl{\rm{ApJ}}                
\def\apjs{\rm{ApJS}}               
\def\aap{\rm{A{\&}A}}              
\def\mnras{\rm{MNRAS}}             

\def\cd{c$\!\!\!\hskip 0.75pt$\raise 0.2pt \hbox{\symbol{24}}}

\def\Msol{\ifmmode{\rm M}_{\mathord\odot}\else M$_{\mathord\odot}$\fi}
\def\Mbh{\ifmmode{M_{\rm BH}}\else{$M_{\rm BH}$}\fi}
\def\REdd{\ifmmode{{\cal R}_{\rm Edd}}\else{${\cal R}_{\rm Edd}$}\fi}
\def\nLn{\ifmmode{$\nu L_{\nu}$}\else{$\nu L_{\nu}$}\fi}

\def\ls{\lower 2pt \hbox{$\;\scriptscriptstyle \buildrel<\over\sim\;$}} 
\def\gs{\lower 2pt \hbox{$\;\scriptscriptstyle \buildrel>\over\sim\;$}}

\def\sed{SED}
\def\seds{SEDs}

\def\kms{\ifmmode{~{\rm km~s^{-1}}}\else{~km~s$^{-1}$}\fi}
\def\ergs{\ifmmode{~{\rm erg~s^{-1}}}\else{~erg~s$^{-1}$}\fi}
\def\m#1{\ifmmode{^{-#1}}\else{$^{-#1}$}\fi}
\def\asec{\ifmmode{^{\prime\prime}}\else{$^{\prime\prime}$}\fi}
\def\asecb{\ifmmode{^{\prime\prime\!\!\!}}\else{$^{\prime\prime\!\!\!}$}\fi}
\def\asecp{\ifmmode{^{\prime\prime\!\!\!}.}\else{$^{\prime\prime\!\!\!}$.}\fi}
\def\deg{\ifmmode{^{\circ}}\else{$^{\circ}$}\fi}
\def\degp{\ifmmode{^{\circ\!\!\!}.}\else{$^{\circ\!\!\!}$.}\fi}
\def\aox{\ifmmode{\alpha_{\rm ox}}\else{$\alpha_{\rm ox}$}\fi}
\def\zbol{\ifmmode{\kappa_{\rm 2-10\; keV}}\else{$\kappa_{\rm 2-10\; keV}$}\fi}

\def\ten#1{\ifmmode{\times 10^{#1}}\else{$\times 10^{#1}$}\fi}
\def\tten#1{\ifmmode{\times 10^{#1}}\else{$\times 10^{#1}$}\fi}
\def\nten#1#2{\ifmmode{#1\times 10^{#2}}\else{$#1\times 10^{#2}$}\fi}

\newcounter{species}
\def\ion#1#2{\setcounter{species}{#2}#1$\;${\sc\roman{species}}\relax}

\def\flion#1#2#3{[{\setcounter{species}{#2}#1$\;${\sc\roman{species}}]$\;\lambda${#3}}\relax}

\def\a{$\alpha$}
\def\b{$\beta$}

\def\pone{paper I}

\def\chandra{{\it Chandra}}

\def\hst{{\it HST}}

\begin{document}

\title{An Assessment of the Energy Budgets of Low-Ionization Nuclear
Emission Regions}

\author{Michael Eracleous\altaffilmark{1,2,3}, 
        Jason A. Hwang\altaffilmark{1,3}, \& 
        H\'el\`ene M. L. G. Flohic\altaffilmark{1,4}}

\altaffiltext{1}{Department of Astronomy and Astrophysics, The Pennsylvania
State University, 525 Davey Lab, University Park, PA 16802} 

\altaffiltext{2}{Center for Gravitational Wave Physics, The Pennsylvania
State University, 104 Davey Lab, University Park, PA 16802}

\altaffiltext{3}{Department of Physics \& Astronomy, Northwestern
University, 2131 Tech Drive, Evanston, IL 60208}

\altaffiltext{4}{Department of Physics and Astronomy, University of
  California, 4129 Frederick Reines Hall, Irvine, CA 92697}

\begin{abstract}
Using the spectral energy distributions (\seds) of the weak active
galactic nuclei (AGNs) in 35 low-ionization nuclear emission regions
(LINERs) presented in a companion paper, we assess whether
photoionization by the weak AGN can power the emission-line
luminosities measured through the large (few-arcsecond) apertures used
in ground-based spectroscopic surveys. Spectra taken through such
apertures are used to define LINERs as a class and constrain
non-stellar photoionization models for LINERs. Therefore, our energy
budget test is a self-consistency check of the idea that the observed
emission lines are powered by an AGN.  We determine the ionizing
luminosities and photon rates by integrating the observed spectral
energy distributions and by scaling a template \sed. We find that even
if all ionizing photons are absorbed by the line-emitting gas, more
than half of the LINERs in this sample suffer from a deficit of
ionizing photons. In 1/3 of LINERs the deficit is severe. If only 10\%
of the ionizing photons are absorbed by the gas, there is an ionizing
photon deficit in 85\% of LINERs. We disfavor the possibility that
additional electromagnetic power, either obscured or emitted in the
unobservable far-UV band, is available from the AGN. Therefore, we
consider other power sources such as mechanical heating by compact
jets from the AGN and photoionization by either young or old stars.
Photoionization by young stars may be important in a small fraction of
cases. Mechanical heating can provide enough power in most cases but
it is not clear how this power would be transferred to the
emission-line gas. Photoionization by post-AGB stars is an important
power source; it provides more ionizing photons that the AGN in more
than half of the LINERs and enough ionizing photons to power the
emission lines in 1/3 of the LINERs. It appears likely that the
emission-line spectra of LINERs obtained from the ground include the
sum of emission from different regions where different power sources
dominate.
\end{abstract}


\section{Introduction}\label{S:intro}

The nature of low-ionization nuclear emission regions (LINERs) has
been the subject of considerable debate since their identification as
a class by \citet{heckman80}. Their defining characteristics are the
relative intensities of their oxygen emission lines, namely
\flion{O}{2}{3727}$\;/\;$\flion{O}{3}{5007}$\;> 1$ and
\flion{O}{1}{6300}$\;/\;$\flion{O}{3}{5007}$\;> 1/3$. They are found
in approximately 50\% of nearby galaxies \citep*{ho97a}, which
suggests that they are an important component of galactic nuclei in
the local universe. The emission lines could be powered by (a)
photoionization by an accreting supermassive black hole, i.e. a
``little monster'' or active galactic nucleus \citep[an AGN;
  e.g.,][]{halpern83, ferland83}, (b) hot stars \citep[either young
  stars in a compact starburst or the hot, exposed cores of evolved
  stars;
  e.g.,][]{terlevich85,filippenko92,shields92,barth00,binette94}, or
(c) shocks \citep[e.g.,][and references therein]{heckman80,
  dopita96a}.  Recent radio, UV, and X-ray surveys at high spatial
resolution and UV variability studies find ``little monsters'' in the
majority of LINERs observed \citep*[e.g.,][]{nagar02, filho04,
  nagar05,filho06, barth98, maoz95, ho01, terashima03, dudik05,
  flohic06, gonzalez06, maoz05}. Thus, LINERs are potentially the
largest subset of AGNs, tracing the AGN population at the lowest
luminosities and providing us an opportunity to study the properties
of the accretion flow in these low-luminosity AGNs.

However, a separate and equally important question remains unanswered:
how important is photoionization by the ``little monsters'' in
powering the emission-line luminosity measured in the spectra of these
objects?  If ``little-monsters'' are not significant or universal
power sources, then we must find alternate sources of power and
understand their relation, if any, to the ``little monster.''  LINERs
and related objects are typically identified in spectroscopic surveys
using telescopes on the ground. Thus, the spectra are taken through
apertures whose characteristic angular size is a few arc-seconds. A
very important survey is that of \citet[][see also references
  therein]{ho97a}, which has systematically identified a large number
of LINERs among nearby, bright galaxies in the northern sky. The
spectra for this survey were taken through a
2\asec$\times$4\asec\ slit, corresponding to a physical dimension of a
few hundred parsec at the distances of the target galaxies.  These
spectra provide some of the main observational constraints for models
that attempt to explain LINERs, specifically, photoionization models
assuming that the ionizing photons come from a ``little
monster''. Therefore, the question we address here is whether the
``little monsters'' can provide enough power and enough ionizing
photons to account for the emission-line luminosity observed through
apertures of a few arc-seconds. This comparison is appropriate because
these are the same apertures through which the defining line ratios of
LINERs and related objects are measured.  In other words, our goal is
to carry out a self-consistency test of non-stellar photoionization
models since these models should not only explain the line ratios but
also the line luminosities.

The question of the energy budget of LINERs has been tackled by a
number of authors in the recent literature. In their evaluation of the
excitation mechanism in a sample of 13 LINERs, \citet*{ho93} found a
deficit of ionizing photons from the AGN, even though the
corresponding photoionization models were successful in reproducing
the relative intensities of the optical emission lines.  They proposed
that either the AGN provides more ionizing photons than they assumed
(which are either extinguished or unobservable) or that some other
power source makes up the deficit.  \citet{maoz95}, using their
near-UV photometry of a sample of 5 UV-bright LINERs and an assumed
shape for the spectral energy distribution (\sed), found that the AGN
could provide enough photons to power the observed emission lines. In
a followup study employing UV spectra of 7 objects, \citet{maoz98}
re-evaluated the energy budgets of LINERs using both AGN and starburst
models for the \sed. They concluded that there is indeed a deficit of
ionizing photons for about half of the objects in their sample,
specifically those with AGN-like UV spectra. They suggested that
extreme-UV/soft-X-ray photons emitted by the AGN provide the missing
power. More recently, \citet{flohic06} assessed the budget of ionizing
photons from the AGN in a sample of 19 LINERs observed in the X-ray
band with \chandra. They extrapolated the observed X-ray power-law to
the Lyman limit and found that the AGN could not provide the requisite
number of ionizing photons. They proposed that either photons from
post-AGB stars from an intermediate-age stellar population or
mechanical power from a jet emanating from the AGN make up the
deficit.

In this paper we re-evaluate the electromagnetic energy budgets of the
``little monsters'' in LINERs using an X-ray selected sample of 35
objects with published data.  In a companion paper \citep*[][hereafter
  \pone]{eracleous09} we put together the \seds\ of these objects and
use a subset of those data to address the question of whether {\it
  photoionization from the weak AGN} can power the (narrow)
emission-line luminosity observed through a
2\asec$\times$4\asec\ aperture in the survey of \citet{ho97b}.  In
\S\ref{S:sample} we present the sample of galaxies and their basic
properties, and we briefly summarize how we collected the data
(details are given in \pone) and how we applied extinction
corrections. In \S\ref{S:ionizing} we describe how we integrated the
\seds\ to compute the ionizing luminosity and rate of ionizing photons
from the AGN. In \S\ref{S:comparison} we compare the ionizing
luminosities and photon rates with the emission-line luminosity and
photon rate, and we find a general deficit of ionizing photons. We
discuss possible ways of balancing the budget in \S\ref{S:discussion}
by asking whether our estimates have missed part of the ionizing
photon output of the AGN and by considering other processes that may
contribute to the excitation of the nebula.

\section{Sample, Spectral Energy Distributions, and Reddening Corrections}\label{S:sample}

\begin{figure}
\centerline{\includegraphics[scale=0.44,angle=0]{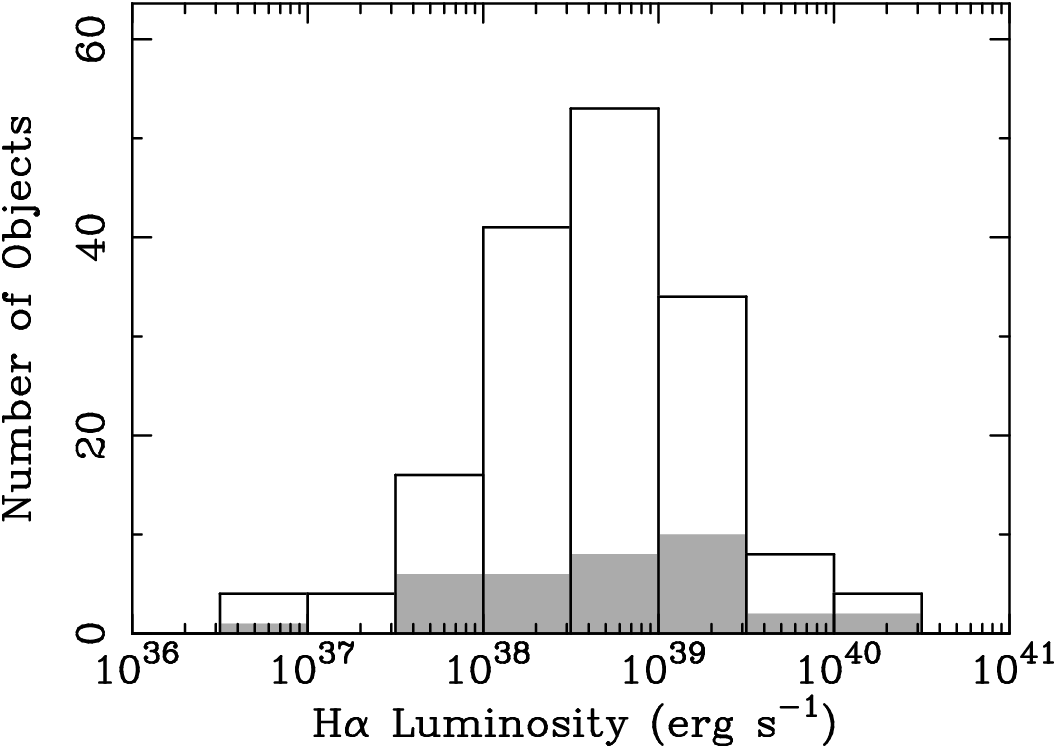}}
\caption{The distribution of H\a\ luminosities of objects in our
sample, compared to that of objects in the \citet{ho97b} ``parent'' sample.
Included in this histogram are LINERs with and without broad emission
lines as well as ``transition objects''. The open bins show the
distribution of H\a\ luminosities of all 164 objects in the
\citet{ho97b} sample plus NGC~1097 and NGC~1553, while the shaded
bins identify the subset of 35 objects included in our sample.
\label{F:histHa}}
\end{figure}

Our sample of LINERs consists of 35 relatively nearby objects (all but
one are at distances less than 40~Mpc) spanning all host morphological
types and with a fairly representative distribution of LINER types.
These are listed in Table~\ref{T:master} along with their basic
properties including LINER types and nuclear H\a\ luminosities from
\citet{ho97b}. The H\a\ luminosities were measured through a
2\asec$\times$4\asec\ slit \citep[see][]{ho95,ho97b} and have
uncertainties of 10--30\% in 27 objects, 30--50\% in one object
(NGC~3608), and 100\% in one object (NGC~3379). In one other object
(NGC~4494), the quoted H-alpha luminosity is a ``$3\sigma$'' upper
limit, while in the 5 remaining objects (NGC~2681, NGC~3031, NGC~4125,
NGC~4736, and NGC~5055) the H\a\ luminosities are lower limits because
the observations were made under non-photometric conditions.  In
Figure~\ref{F:histHa} we compare the distribution of
H\a\ luminosities among the objects in our sample with that of similar
objects in the \cite{ho97b} ``parent'' sample (i.e., LINERs with and
without broad lines and ``transition objects''). As this figure shows,
our objects span the entire range of luminosities of the \cite{ho97b}
parent sample, although high-luminosity objects are somewhat
over-represented.

\def\od{\phantom{0}}
\def\td{\phantom{00}}
\begin{deluxetable*}{llrrrrrc}
\tabletypesize{\scriptsize}
\tablewidth{5.5in}
\tablecaption{Sample of LINERs and Their Properties\label{T:master}}
\tablehead{
\colhead{} &
\colhead{} &
\colhead{} &
\multicolumn{2}{c}{Measured \sed} &
\multicolumn{2}{c}{Quasar \sed} \\
\noalign{\vskip -3 pt}
\colhead{} &
\colhead{LINER} &
\colhead{} &
\multicolumn{2}{c}{\hrulefill} &
\multicolumn{2}{c}{\hrulefill} \\
\colhead{Galaxy\phantom{(M84, 3C 272.1)}} &
\colhead{Type \tablenotemark{\rm a}} &
\colhead{$\log L_{\rm H\alpha}\;$\tablenotemark{{\rm a,b}}} &
\colhead{$\log L_{\rm i}\;$\tablenotemark{\rm b}} &
\colhead{$\log Q_{\rm i}\;$\tablenotemark{\rm b}} &
\colhead{$\log L_{\rm i}\;$\tablenotemark{\rm b}} &
\colhead{$\log Q_{\rm i}\;$\tablenotemark{\rm b}} &
\colhead{$\log Q_{\star,m_{44}}\;$\tablenotemark{\rm b,d}} \\
\noalign{\vskip 3 pt}
\colhead{(1)\phantom{(M84, 3C 272.1)}} &
\colhead{(2)} &
\colhead{(3)} &
\colhead{(4)} &
\colhead{(5)} &
\colhead{(6)} &
\colhead{(7)} &
\colhead{(8)} 
}
%
%
\startdata
  NGC 0266                 & L1      & $ 39.30$ & $<41.81$ & $<51.59$ & $ 41.86$ & $ 52.01$ & 50.99  \\ 
  NGC 0404                 & L2      & $ 37.63$ & $ 38.94$ & $ 49.37$ & $ 37.57$ & $ 47.72$ & 48.69  \\
  NGC 1097                 & L1      & $ 38.72$ & $ 41.52$ & $ 51.43$ & $ 41.62$ & $ 51.77$ & \dots  \\
  NGC 1553                 & L2/T2   & $ 39.30$ & $ 40.99$ & $ 50.72$ & $ 40.92$ & $ 51.07$ & \dots  \\
  NGC 2681                 & L1      & $>38.83$ & $<39.35$ & $<49.52$ & $ 39.24$ & $ 49.39$ & \dots  \\
  NGC 3031 (M81)           & S1.5/L1 & $>38.46$ & $ 40.97$ & $ 50.55$ & $ 41.26$ & $ 51.41$ & \dots  \\
  NGC 3169                 & L2      & $ 39.02$ & $<42.80$ & $<52.98$ & $ 42.03$ & $ 52.18$ & 50.45  \\
  NGC 3226                 & L1      & $ 38.93$ & $<41.90$ & $<52.09$ & $ 41.68$ & $ 51.83$ & 50.18  \\
  NGC 3379 (M105)          & L2/T2   & $ 37.94$ & $<38.25$ & $<48.37$ & $ 38.21$ & $ 48.36$ & 50.13  \\
  NGC 3507                 & L2      & $ 39.39$ & $<38.61$ & $<48.74$ & $<38.58$ & $<48.73$ & 50.32  \\
  NGC 3607                 & L2      & $ 38.93$ & $<38.34$ & $<47.42$ & $<38.68$ & $<48.83$ & 50.44  \\
  NGC 3608                 & L2/S2   & $ 38.28$ & $<39.79$ & $<49.92$ & $<39.76$ & $<49.91$ & 50.64  \\
  NGC 3628                 & T2      & $ 36.87$ & $<37.72$ & $<47.84$ & $<37.68$ & $<47.83$ & 48.03  \\
  NGC 3998                 & L1      & $ 40.00$ & $ 42.36$ & $ 52.40$ & $ 42.40$ & $ 52.55$ & 50.87  \\
  NGC 4111                 & L2      & $ 39.40$ & $<40.30$ & $<49.90$ & $<40.54$ & $<50.69$ & 50.74  \\
  NGC 4125                 & T2      & $>38.96$ & $<39.75$ & $<49.87$ & $<39.72$ & $<49.87$ & \dots  \\
  NGC 4143                 & L1      & $ 38.69$ & $<41.02$ & $<51.10$ & $ 41.01$ & $ 51.17$ & 50.41  \\
  NGC 4261 (3C 270)        & L2      & $ 39.35$ & $<42.08$ & $<52.13$ & $ 42.00$ & $ 52.15$ & 50.83  \\
  NGC 4278                 & L1      & $ 39.17$ & $<40.95$ & $<51.01$ & $ 40.94$ & $ 51.09$ & 50.11  \\
  NGC 4314                 & L2      & $ 38.45$ & $<39.58$ & $<50.03$ & $<38.47$ & $<48.62$ & 49.84  \\
  NGC 4374 (M84, 3C 272.1) & L2      & $ 38.89$ & $ 40.31$ & $ 50.05$ & $ 40.53$ & $ 50.68$ & 50.33  \\
  NGC 4438                 & L1      & $ 39.37$ & $<39.77$ & $<49.23$ & $<40.06$ & $<50.21$ & 49.97  \\
  NGC 4457                 & L2      & $ 39.57$ & $<40.07$ & $<50.21$ & $ 39.99$ & $ 50.14$ & 50.83  \\
  NGC 4486 (M87, 3C 274)   & L2      & $ 39.44$ & $ 41.34$ & $ 51.51$ & $ 41.18$ & $ 51.33$ & 50.07  \\
  NGC 4494                 & L2      & $ 37.54$ & $<39.99$ & $<50.11$ & $ 39.95$ & $ 50.10$ & 50.03  \\
  NGC 4548 (M91)           & L2      & $ 38.46$ & $<40.41$ & $<49.50$ & $ 40.72$ & $ 50.87$ & 49.88  \\
  NGC 4552 (M89)           & T2      & $ 38.52$ & $ 40.36$ & $ 50.20$ & $ 40.40$ & $ 50.55$ & 50.44  \\
  NGC 4579 (M58)           & L1      & $ 39.44$ & $ 41.91$ & $ 51.26$ & $ 42.23$ & $ 52.38$ & 50.37  \\
  NGC 4594 (M104)          & L2      & $<39.70$ & $ 41.42$ & $ 51.67$ & $ 40.86$ & $ 51.01$ & 50.94  \\
  NGC 4636                 & L1      & $ 38.27$ & $<39.57$ & $<49.34$ & $<39.77$ & $<49.92$ & 49.95  \\
  NGC 4736 (M94)           & L2      & $>37.75$ & $ 39.72$ & $ 49.71$ & $ 39.76$ & $ 49.91$ & \dots  \\
  NGC 5055 (M63)           & T2      & $>37.91$ & $ 41.81$ & $ 52.34$ & $ 39.28$ & $ 49.43$ & \dots  \\
  NGC 5866                 & T2      & $ 38.01$ & $<39.18$ & $<48.55$ & $<39.48$ & $<49.63$ & 49.71  \\
  NGC 6500                 & L2      & $ 40.31$ & $ 41.92$ & $ 52.25$ & $ 40.71$ & $ 50.86$ & 50.68  \\
  NGC 7331                 & T2      & $ 38.49$ & $<38.43$ & $<48.41$ & $<38.52$ & $<48.67$ & 50.21  \\
\enddata
\tablenotetext{a}{The LINER types and H\a\ line luminosities (measured 
                  through a 2\asec$\times$4\asec aperture) were taken
                  from \citet*{ho97b}, with the exception of NGC~1097
                  and NGC~1553. The LINER types for these two galaxies
                  were taken from \citet{phillips84}
                  and \citet{phillips86} respectively. The
                  emission-line luminosity of NGC~1097 was taken
                  from \citet{storchi93}, while that of NGC~1553 was
                  taken from \citet{phillips86}. L1=LINER with broad a
                  H$\alpha$ line, L2=LINER without broad a H$\alpha$
                  line, T2=intermediate emission line ratios between
                  LINER and \ion{H}{2} region, S=Seyfert, combinations
                  indicate intermediate line ratios between two
                  classes. The uncertainty on the H\a\ luminosity is
                  10--30\%, with the exception of NGC~3608 (30--50\%)
                  and NGC~3379 (100\%). In the case of NGC~4494, we
                  list a ``3$\sigma$'' upper limit on the H\a\
                  luminosity. All upper limits on the H\a\ luminosity
                  are the result of non-photometric sky conditions.}
\tablenotetext{b}{The luminosities are measured in \ergs and the photon rates 
                  in s$^{-1}$.}
\tablenotetext{c}{\REdd\ is the Eddington ratio, defined as the ratio of 
                  the bolometric luminosity to the Eddington
                  luminosity.}
\tablenotetext{d}{The ionizing photon rate from post-AGB stars in 
                  the old stellar population.  It is estimated from
                  the spectroscopic B magnitude reported
                  by \citet{ho97b} after converting it to a standard B 
                  magnitude and using the prescription
                  of \citet{binette94}. See details and discussion
                  in \S\ref{S:discussion} of the text.}

\end{deluxetable*}

The sample galaxies were observed in the X-ray band with the \chandra\
X-Ray Observatory with moderately long exposure times. In 24 objects
the X-ray emission from a weak AGN was detected by \chandra, while for
the remaining 11 objects an upper limit was obtained. We used data
from the literature to construct the spectral energy distributions
(\seds) of these weak AGNs from the radio to the X-ray bands, which we
present in \pone\ along with derived quantities such as the bolometric
luminosity, the Eddington ratio, and the optical-to-X-ray spectral
index, \aox\footnote{The optical-to-X-ray spectral index is the
exponent of a power-law connecting the monochromatic luminosity
densities at 2500~\AA\ and 2~keV. If $L_{\nu}\propto\nu^{-\aox}$,
\aox\ is given by  $\aox = 1+0.384\; \log\left[(\nu
L_{\nu})_{\rm 2500\;\AA}/(\nu L_{\nu})_{\rm 2\;keV}\right]$.}.
Included in \pone\ are also references to the sources of the data as
well as an extensive discussion of relevant caveats.  The 2--10~keV
X-ray luminosities of the detected AGNs are between $1.7\times
10^{38}$ and $2.6\times 10^{41}~\ergs$, while the undetected AGNs have
$L_{\rm 2-10\;keV} < 1.2\times 10^{39}~\ergs$.  Their bolometric
luminosities were derived by integrating the observed \sed\ in a small
fraction of objects or by scaling the X-ray luminosity for the
majority of objects. Detected AGNs in this sample have $1.2\times
10^{38}~\ergs < L_{\rm bol} < 3.3\times 10^{43}~\ergs$ while the
undetected objects have $L_{\rm bol} < 1.1\times 10^{41}~\ergs$ (see
the detailed discussion in \S4.1 of \pone). Using black hole masses
obtained from the stellar velocity dispersion in the host galaxies, we
estimated the Eddington ratios (defined as the ratio of the bolometric
luminosity to the Eddington luminosity; see \S2 and \S4.1 of
\pone). We find that $3\times 10^{-8} < \REdd < 4\times 10^{-4}$ for
X-ray detected AGNs, while undetected AGNs have $\REdd< 2\times
10^{-5}$ (see histogram in Fig.~5 of \pone).

Here we are primarily interested in the ionizing portion of the
\sed\ at $E>1\;{\rm Ry}$. To define the \sed\ at these energies we
relied on UV photometry and spectroscopy of the nuclei carried out
with the {\it Hubble Space Telescope} (\hst) and X-ray observations
carried out with \chandra\ (see references in \pone). High spatial
resolution in the UV and X-ray bands is essential in order to separate
the AGN from discrete and diffuse sources in its immediate vicinity.
From the UV observations we have obtained the monochromatic luminosity
density at 2500~\AA\ for 19 objects. We have used UV variability
information \citep[from][]{maoz05} and UV spectra \citep[primarily
  from][]{maoz98} to separate ``little monsters'' from compact
starbursts. By studying the distribution of \aox\ of these two types
of nuclei in \pone, we found that ``little monsters'' always have
$\aox < 1.5$.  Adopting $\aox < 1.5$, we derived a {\it generous}
upper limit for the 2500~\AA\ fluxes of AGNs in our sample that were
observed in the X-ray, but not in the UV band. We consider such UV
upper limits rather generous because our adopted limit on
\aox\ corresponds to quasars and AGNs that are much more luminous than
the objects in our sample \citep[$|\aox|$ decreases with decreasing UV
  luminosity, cf][]{strateva05,steffen06}\footnote{It is noteworthy
  that \aox\ may have some dependence on redshift, as argued by
  \citet{kelly07}. Nevertheless, when the dependence on redshift is
  taken into account, the value we have adopted as an upper limit on
  \aox\ still corresponds to quasars and AGNs that are much more
  luminous than the objects of our sample. It is also noteworthy that
  \citet*{tang07} have argued that a correlation between \aox\ and UV
  luminosity is not firmly established, although they do conclude that
  such a correlation is likely present in a complete subsample of the
  \cite{steffen06} data.}.

In an effort to recover the intrinsic luminosities of the AGNs we
corrected their UV and X-ray fluxes for extinction. In the X-ray band,
extinction corrections are more reliable than in the UV since the
column density can be determined directly by fitting the observed
X-ray spectrum (see details in \pone).  Extinction corrections in the
UV, involve a number of assumptions, and as a result they are subject
to considerable uncertainty. Therefore, we have tried to err on the
side of caution and overestimate the intrinsic luminosity rather than
underestimating it. To obtain the total reddening towards the UV
source, we used the total hydrogen column densities derived from fits
of an absorbed power-law model to the X-ray spectra. We converted
these to a color excess using the relation between column density and
visual extinction for the Milky Way, $N_{\rm H}/A_{\rm V} = 1.79\times
10^{21}~\rm{cm^{-21}~mag^{-1}}$ \citep{predhl95}, and the
\citet{seaton79} extinction law, which implies that $R_{\rm V}\equiv
A_{\rm V}/E(B-V)=3.2$. The resulting values of $E(B-V)$ are given in
\pone\ along with the corresponding values for the interstellar medium
of the Milky Way in the direction of the source \citep*[taken
  from][]{schlegel98}.  Comparing the two values of the color excess,
we find that the value derived from the X-ray spectrum is comparable
to or greater than the Milky Way value, suggesting that the former
value also captures extinction in the host galaxy of the AGN. Using
the color excess derived from the X-ray spectrum we computed
extinction corrections for the UV data by separately applying the
Milky Way extinction laws of \citet{seaton79} and \citet*{cardelli89},
the Large and Small Magellanic Cloud extinction laws of
\citet{korneef81} and \citet{bouchet85}, respectively, and the
starburst galaxy extinction law of \citet*{calzetti94}. In the end we
adopted the \citet{seaton79} law since this gives the largest
correction. We note, however, that UV luminosities obtained with the
other Milky Way or Magellanic Cloud laws are less than 10\% lower than
the values we have adopted.  The corrected UV fluxes are included in
\pone\ along with a discussion of the differences between the five
correction schemes (see \S3.2 of \pone).

\section{Ionizing Luminosities and Photon Rates}\label{S:ionizing}

We evaluated the 1~Ry--100~keV ionizing luminosities ($L_{\rm
  1\;Ry-100\;keV}$) and ionizing photon rates ($Q_{\rm
  1\;Ry-100\;keV}$) of the AGNs in our sample, after extrapolating
their 2--10~keV X-ray spectra, in two different ways. We First
integrated the \sed, assuming that pairs of points could be connected
by a power law. Under this assumption, the luminosity and photon rate
of each segment were computed analytically and then the contributions
of different segments were summed, i.e.,
$$
\quad\phantom{\rm and}\quad
L_{\rm 1\;Ry-100\;keV} = \sum_{i=1}^{N-1} 
\int_{\nu_i}^{\nu_{i+1}} L_{\nu}\; d\nu 
$$
\begin{equation}
\quad{\rm and}\quad
Q_{\rm 1\;Ry-100\;keV} = \sum_{i=1}^{N-1} 
\int_{\nu_i}^{\nu_{i+1}} {L_{\nu}\over h\nu}\; d\nu \; .
\label{Q:sum}
\end{equation}
The upper limit of integration is appropriate because the X-ray
spectra of Seyfert galaxies typically cut off at energies between 50
and 150~keV \citep{molina09} \footnote{If we extrapolate the observed
  X-ray spectra to 500~keV instead, the resulting ionizing luminosity
  increases only by a few percent and the ionizing photon rate
  increases by a negligible amount}. The results are listed in columns
(4) and (5) of Table~\ref{T:master}. If at least one of the measured
points in an \sed\ is an upper limit we take $L_{\rm 1\;Ry-100\;keV}$
and $Q_{\rm 1\;Ry-100\;keV}$ to be upper limits as well.

The above scheme relies on the assumption that the ionizing \sed\ of
the AGN is a power law between the near-UV and soft X-ray bands
(typically 2500~\AA\ to 0.5~keV), which is not necessarily
correct. The \seds\ of more luminous AGNs, such as Seyfert galaxies and
quasars, are thought to have a significant flux in the far-UV band
\citep[see, for example,][]{elvis94}, which would not be observable in
the case of the weak AGNs of our sample. To deal with this
possibility we also evaluated the 1~Ry--100~keV ionizing luminosities
and ionizing photon rates of the AGNs in our sample by assuming that
their \seds\ are similar to the standard radio-quiet quasar
\sed\ presented by \citet{elvis94}.  Even though the weak AGNs in
LINERs are thought to be {\it radio-loud} \citep[e.g.,][]{nagar01,
  terashima03, ulvestad01, anderson04}, we have opted to use the {\it
  radio-quiet} quasar template because it has a higher value of
\aox\ than the {\it radio-loud} quasar template (1.36 $vs$ 1.34).  In
other words, we are conservatively adopting the higher UV luminosity
for a given X-ray luminosity. We also note that these values of
\aox\ are higher than those of 80\% of the weak AGNs in our sample,
which means that by using them we are overestimating the UV luminosity
and photon rate for the vast majority of these weak AGNs.  The
standard radio-quiet quasar \sed, when normalized to a 2--10~keV X-ray
luminosity of $L_{\rm 2-10\; keV}=1\times 10^{40}~{\rm erg~s}^{-1}$,
yields an ionizing luminosity of $L_{\rm 1\;Ry-100\; keV}=9.77\times
10^{40}~{\rm erg~s}^{-1}$ and an ionizing photon rate of $Q_{\rm
  1\;Ry-100\; keV}=1.38\times 10^{51}~{\rm erg~s}^{-1}$ after
extrapolating the X-ray power-law portion to 100~keV. We obtained the
ionizing luminosities and ionizing photon rates of our sample AGNs by
scaling their 2--10~keV X-ray luminosities accordingly. The results
are listed in columns (6) and (7) of Table~\ref{T:master}. In cases
where we only have an upper limit to the 2--10~keV luminosity we take
$L_{\rm 1\;Ry-100\;keV}$ and $Q_{\rm 1\;Ry-100\;keV}$ to be upper
limits as well. This scheme has some advantages over the previous one:
(i) it bypasses uncertainties in extinction corrections in the UV
since it relies only on the X-ray extinction, which is derived by
fitting the X-ray spectrum, (ii) it depends only on the 2-10~keV X-ray
luminosity, which is the quantity measured for most of our objects,
and (iii) it bypasses indirect determinations of upper limits to the
UV flux through \aox. On the negative side, this scheme is sensitive
to the assumed shape of the \sed. Theoretical models
\citep[e.g.,][]{dimatteo00,ball01,ptak04} suggest that the \seds\ of
weak AGNs in LINERs have a lower far-UV luminosity relative to the
X-ray luminosity than Seyfert galaxies and quasars. On the other hand,
\citet{maoz07} has argued that the values of \aox\ of such weak AGNs
are compatible with the values found among Seyfert galaxies, which
suggests that the \seds\ of weak AGNs in LINERs include the same
number of UV photons per unit X-ray luminosity as those of Seyferts.
Indeed, our study of the average \sed\ of the little monsters in 
\pone\ bolsters this view.

\section{Results: Comparison of Ionizing Power and Emission-Line Power}\label{S:comparison}

\begin{figure*}
\centerline{
\vbox{\hsize=3.truein
\rightline{\includegraphics[scale=0.45,angle=0]{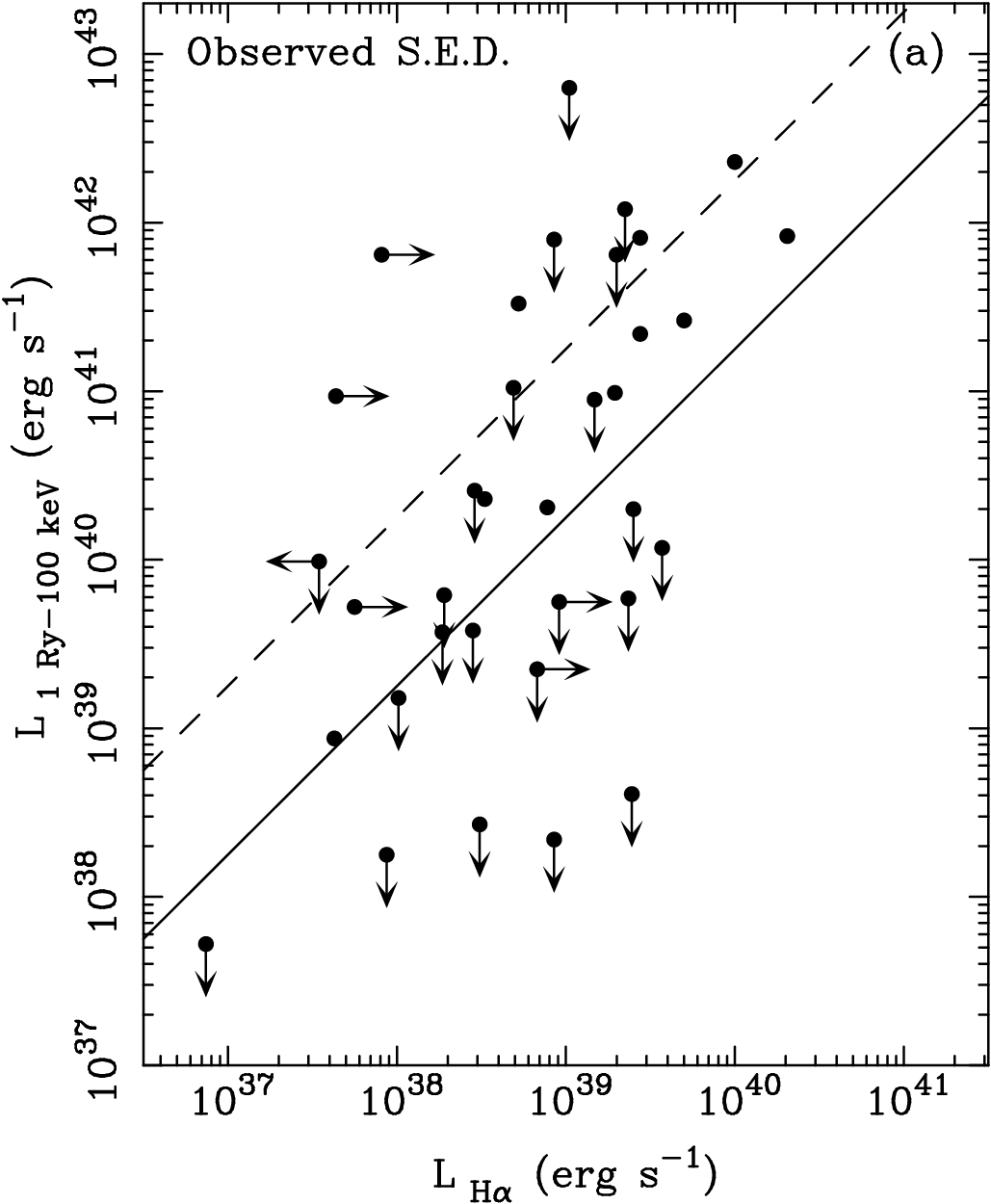}}
\vskip 0.3truein
\rightline{\includegraphics[scale=0.45,angle=0]{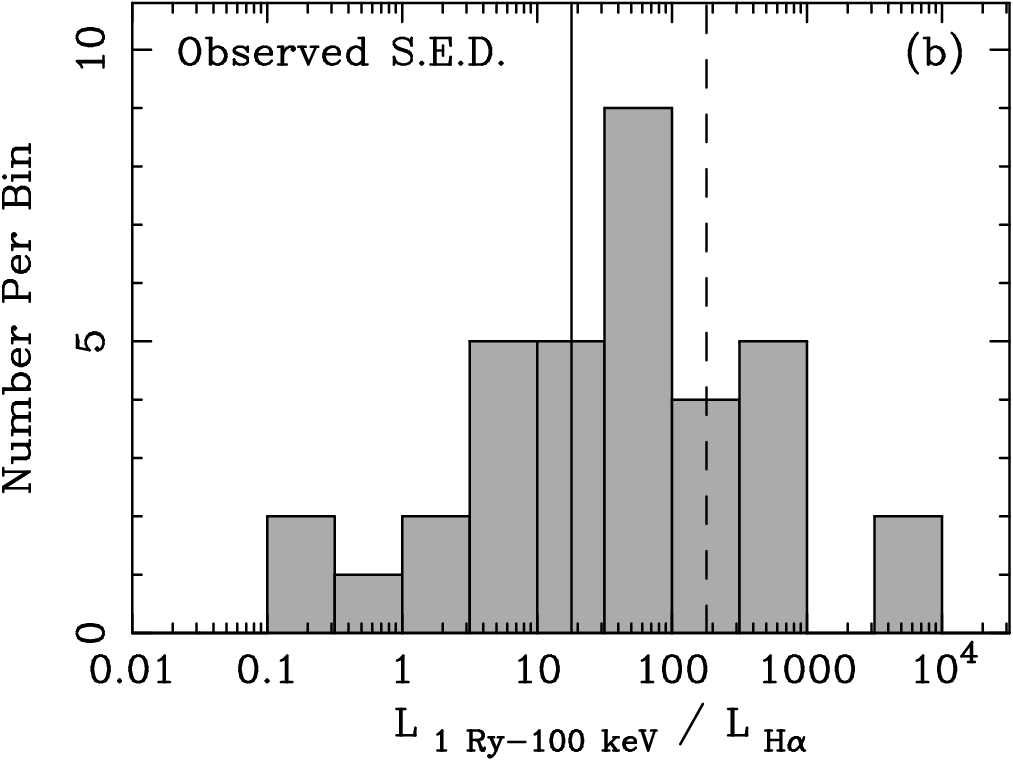}}
}
\hskip 0.3truein 
\vbox{\hsize=3.truein
\rightline{\includegraphics[scale=0.45,angle=0]{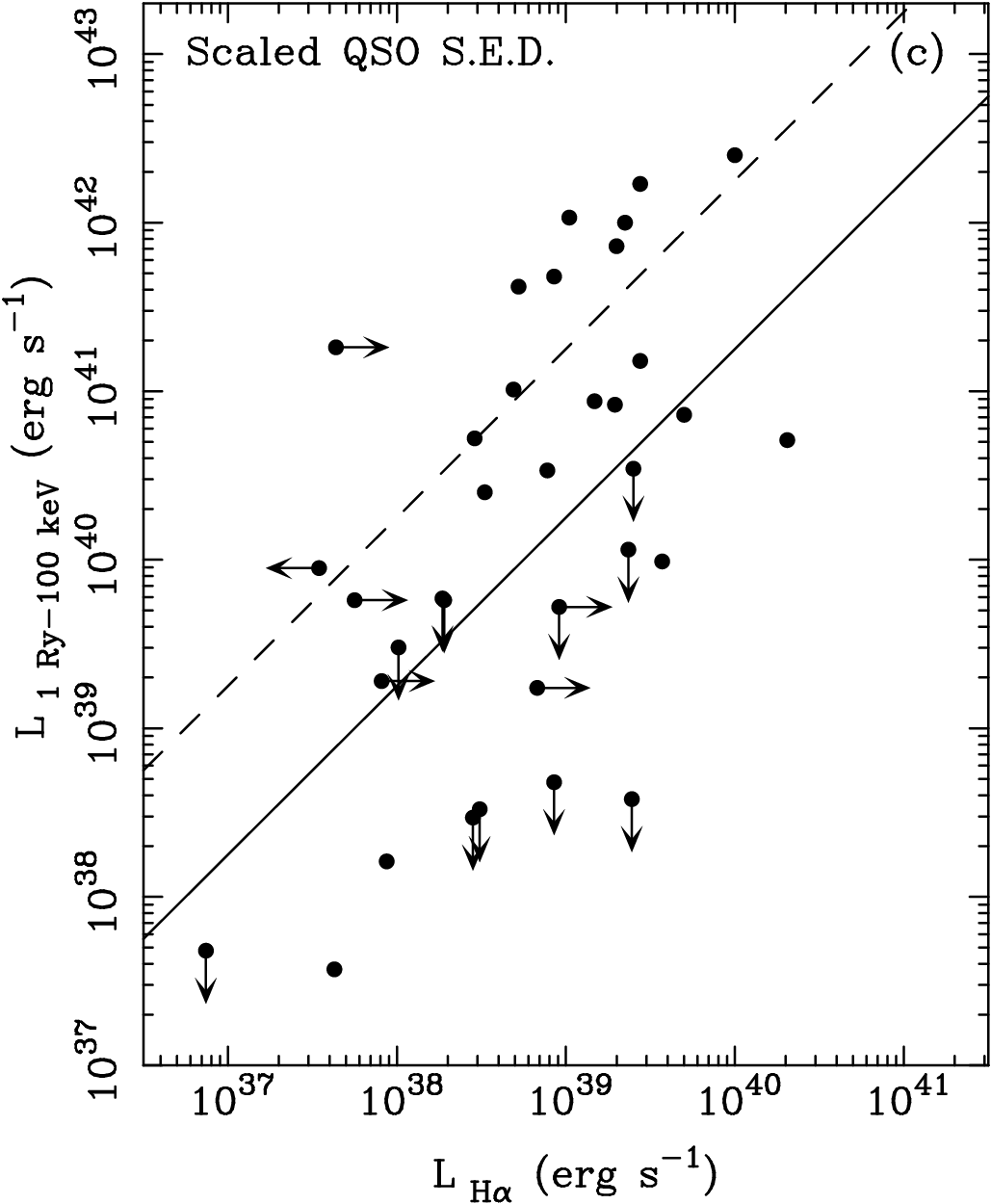}}
\vskip 0.3truein
\rightline{\includegraphics[scale=0.45,angle=0]{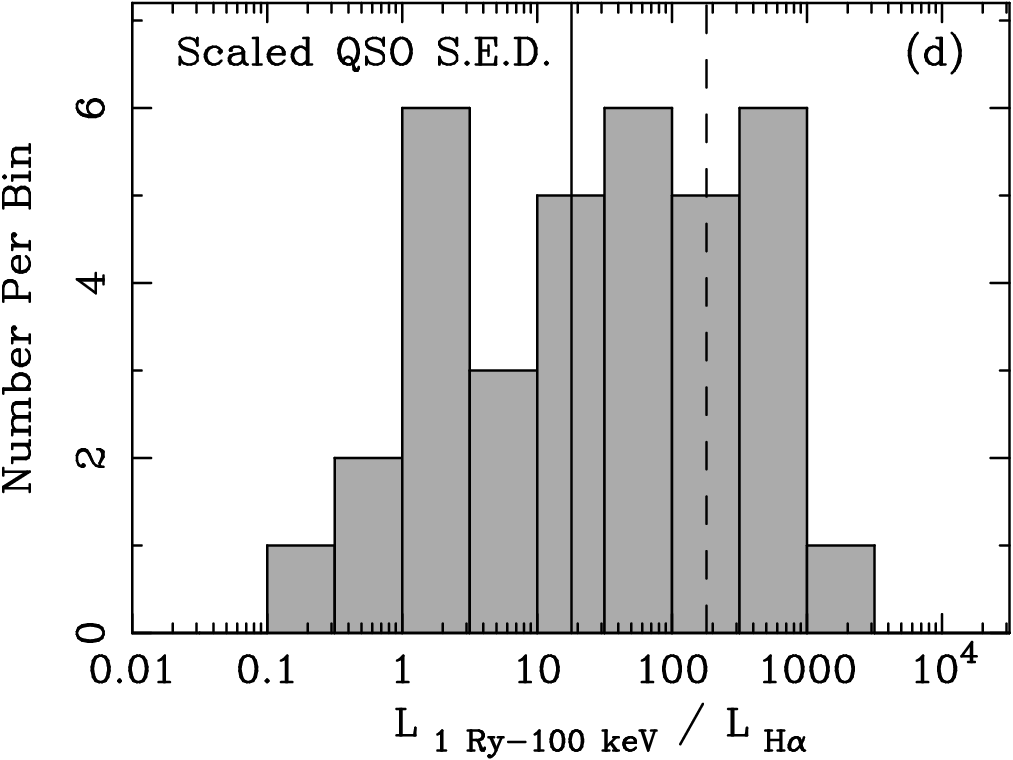}}
}
}
\caption{(a) The ionizing (1~Ry--100~keV) luminosity of LINERs in our
sample, plotted against their H\a\ luminosity. The ionizing luminosity
was determined by directly integrating the \sed, as described in
\S\ref{S:ionizing}. Arrows denote upper limits on the ionizing
luminosity.  The ionizing luminosity balances the cooling rate of a
uniform, photoionized slab of a given H\a\ luminosity along the solid
line ($L_{\rm 1\;Ry-100\;keV}=L_{\rm cool}=18\;L_{\rm H\alpha}$; see
\S\ref{S:comparison} of the text). If only a fraction $f_{\rm
c}$(=0.1) of the ionizing photons are absorbed by the nebula, then
energy balance is represented by the dashed line ($L_{\rm
1\;Ry-100\;keV}=L_{\rm cool}/f_{\rm c}$). (b) An alternative
representation of (a) in the form of a histogram of the ratio $L_{\rm
1\;Ry-100\;keV}/L_{\rm H\alpha}$. The solid and dashed lines have the
same meaning as in (a). (c) Same as (a) but with ionizing luminosities
measured by scaling a standard quasar \sed\ to the 2--10~keV
luminosity of the AGN. (d) Same as (b) but with ionizing luminosities
measured by scaling a standard quasar \sed\ to the 2--10~keV
luminosity of the AGN.
\label{F:LionLHa}}
\end{figure*}

To assess the energy budget of the LINERs in our sample we compare in
Figure~\ref{F:LionLHa}, the ionizing luminosities of their weak AGNs
to the corresponding H\a\ luminosities. In Figure~\ref{F:LionLHa}a we
plot the ionizing luminosity obtained by integrating the AGN
\sed\ against the H\a\ luminosity. In order to use this plot for our
goals, we need to relate the H\a\ luminosity of the emission line gas
to its total cooling rate, which we do by appealing to photoionization
models. In particular, we use the models computed by \citet{lewis03},
employing the photoionization code Cloudy \citep[v94.0;
  see][]{ferland98} and adopting a representative low-luminosity AGN
\sed. These models spanned a wide range of ionization parameters
($\log U=-4.5, -4.0, -3.5, -3.0$), densities ($\log [n/{\rm
    cm}^{-3}]=2, 3, 4, 5, 6$), and metallicities ($Z/Z_{\odot}=0.3, 1,
2$). By re-examining the results of these calculations, we find a
tight relation between the total cooling rate of the nebula and its
H\a\ luminosity, namely $\langle L_{\rm cool}/L_{\rm H\alpha}\rangle =
18$ with a dispersion of 2 about the mean. From this we conclude that
energy balance in the nebula requires that $L_{\rm 1\;Ry-100\;keV} >
18\; L_{\rm H\alpha}$. This condition is represented by a solid line
in Figure~\ref{F:LionLHa}a.  We note that 12/35 LINERs in our sample
do not meet this most fundamental requirement. In practice, it is
quite possible that only a fraction, $f_{\rm c}$, of the ionizing
luminosity of the AGN are absorbed by the line-emitting gas. This
could be because the gas is not completely optically thick to the
ionizing photons, because it is porous (i.e., inhomogeneous on
small scales), or because it has a large-scale geometry such that it
only covers a fraction of the sky, as seen from the AGN (e.g., it
could be toroidal). In such a case, the condition for energy balance
becomes $L_{\rm 1\;Ry-100\;keV} > 18\; L_{\rm H\alpha} / f_{\rm
  c}$. We illustrate an example of this new condition, for $f_{\rm c}
= 0.1$, in Figure~\ref{F:LionLHa}a as a dashed line.  If $f_{\rm c} =
0.1$, 24/35 LINERs in our sample fail this stricter
requirement. Figure~\ref{F:LionLHa}b is an alternative representation
of Figure~\ref{F:LionLHa}a where we show the distribution of the ratio
$L_{\rm 1\;Ry-100\;keV} / L_{\rm H\alpha}$ (it is effectively a
projection of the data along the diagonal lines in
Fig.~\ref{F:LionLHa}a). The solid and dashed lines in
Figure~\ref{F:LionLHa}b have the same meaning as in
Figure~\ref{F:LionLHa}a.

Figures~\ref{F:LionLHa}c,d are analogous to Figures~\ref{F:LionLHa}a,b
with the difference that the ionizing luminosities and photon rates
were obtained by scaling the standard quasar \sed.
Figure~\ref{F:LionLHa}c leads to very similar conclusions as
Figure~\ref{F:LionLHa}a; if we assume that all ionizing photons are
absorbed by the emission-line nebula, then 14/35 LINERs fail the
energy budget test. If we apply the stricter condition that only 10\%
of the ionizing luminosity contributes to the heating of the nebula,
then 24/35 LINERs appear to suffer from a deficit in their energy
budgets.


\begin{figure*}
\centerline{
\vbox{\hsize=3.truein
\rightline{\includegraphics[scale=0.45,angle=0]{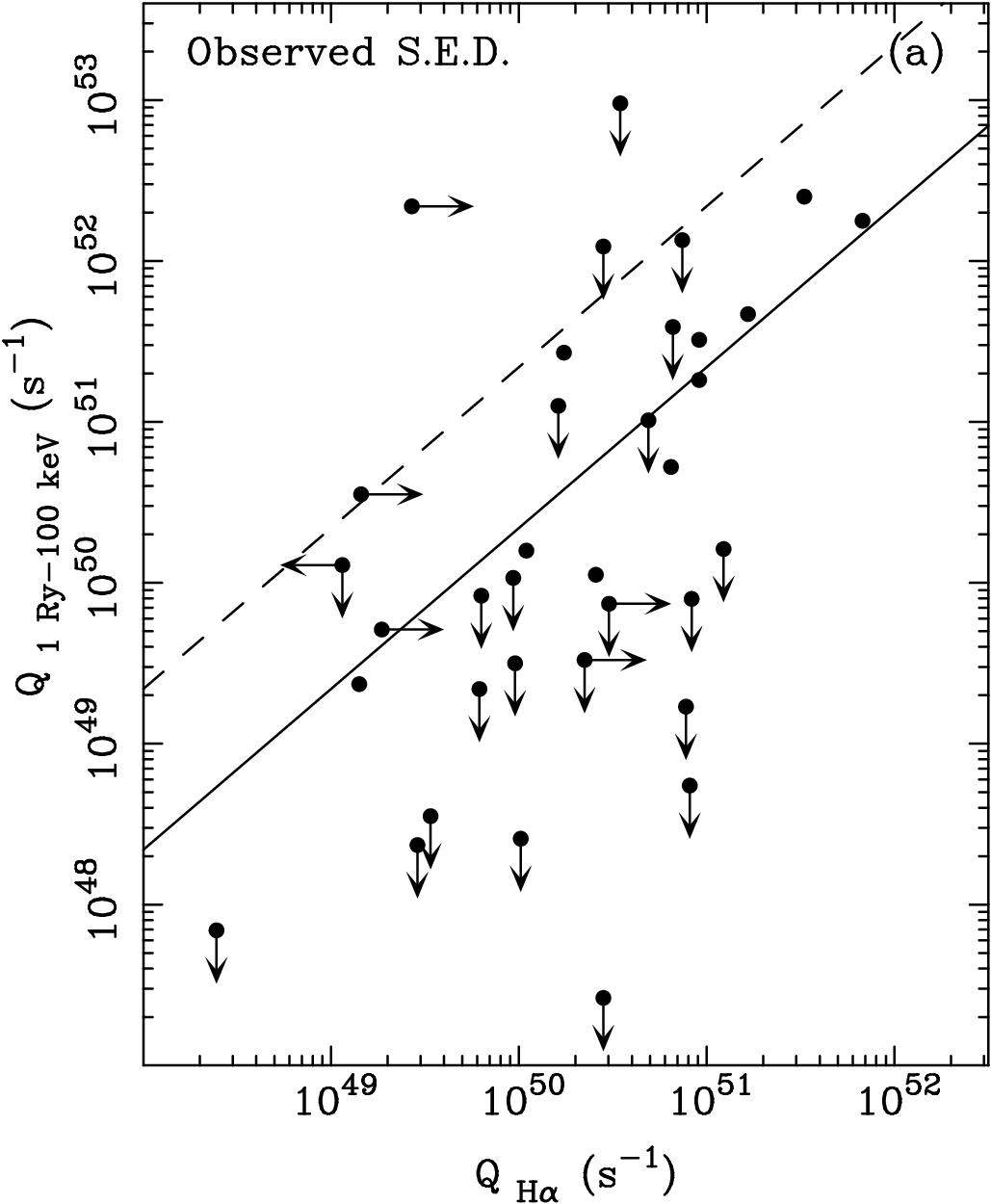}}
\vskip 0.3truein
\rightline{\includegraphics[scale=0.45,angle=0]{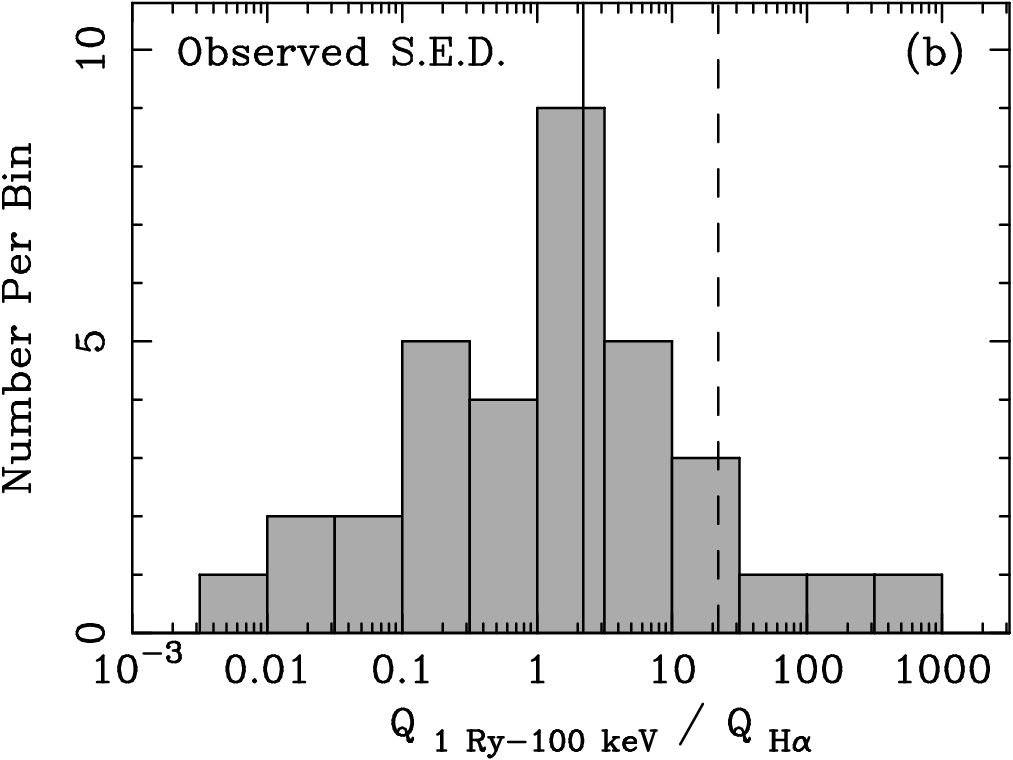}}
}
\hskip 0.3truein 
\vbox{\hsize=3.truein
\rightline{\includegraphics[scale=0.45,angle=0]{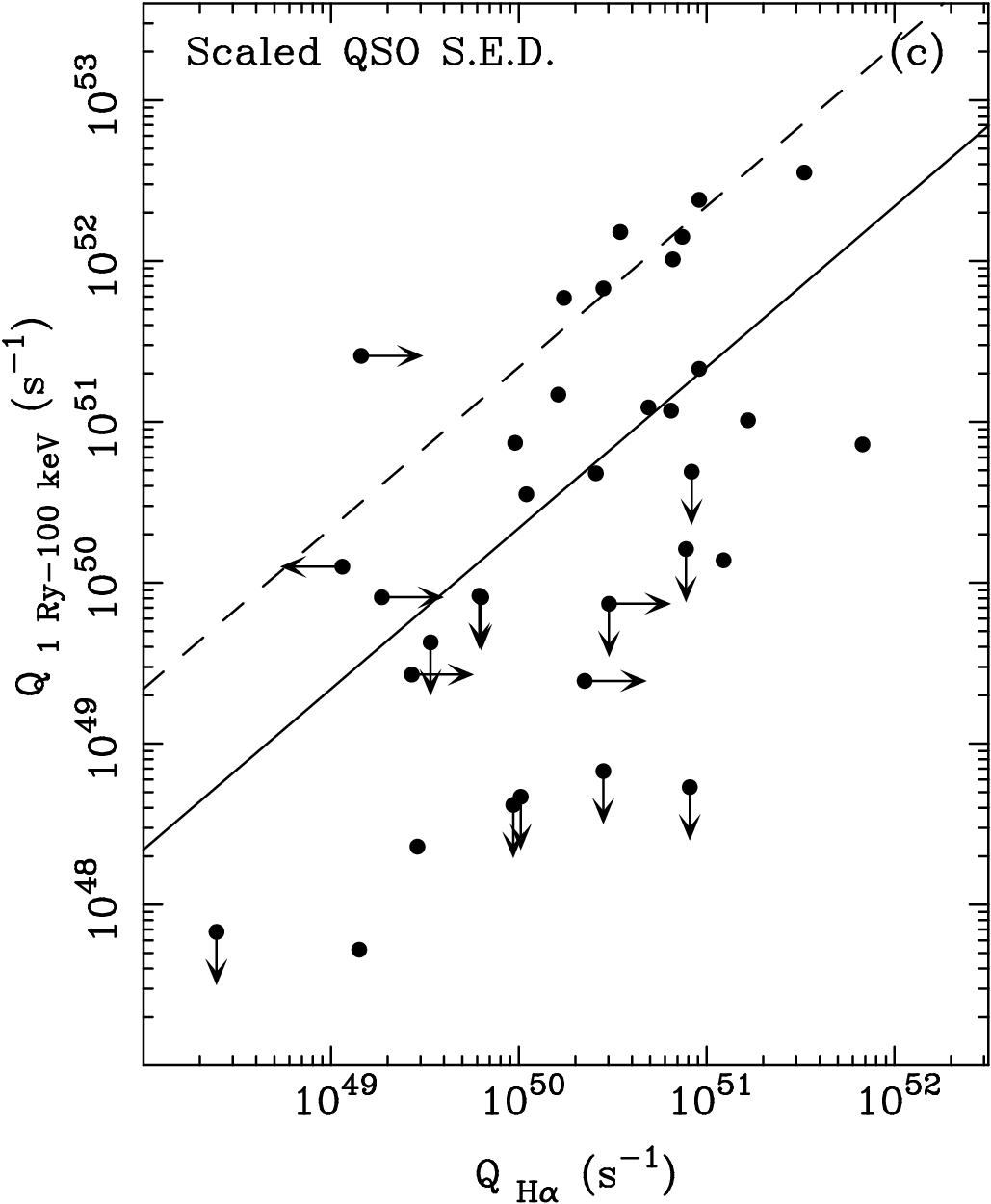}}
\vskip 0.3truein
\rightline{\includegraphics[scale=0.45,angle=0]{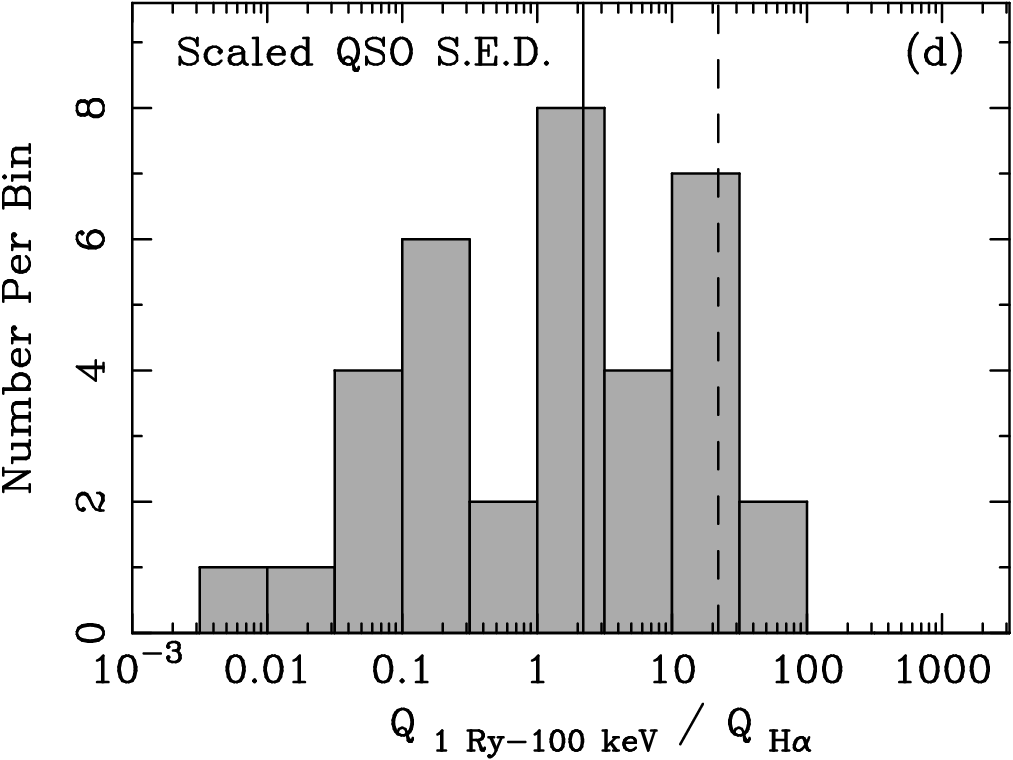}}
}
}
\caption{(a) The ionizing (1~Ry--100~keV) photon rate of LINERs in our
sample, plotted against their H\a\ photon rate. The ionizing photon
rate was determined by directly integrating the \sed, as described in
\S\ref{S:ionizing}. Arrows denote upper limits on the ionizing photon
rate.  The ionizing photon rate balances the H$\alpha$ photon rate
along the solid line ($Q_{\rm 1\;Ry-100\;keV}=2.2\;Q_{\rm H\alpha}$;
see \S\ref{S:comparison} of the text). If only a fraction $f_{\rm
c}$(=0.1) of the ionizing photons are absorbed by the nebula, then
photon number balance is represented by the dashed line ($Q_{\rm
1\;Ry-100\;keV}=2.2\;Q_{\rm H\alpha}/f_{\rm c}$). (b) An alternative
representation of (a) in the form of a histogram of the ratio $Q_{\rm
1\;Ry-100\;keV}/Q_{\rm H\alpha}$. The solid and dashed lines have the
same meaning as in (a). (c) Same as (a) but with ionizing photon rates
measured by scaling a standard quasar \sed\ to the 2--10~keV
luminosity of the AGN. (d) Same as (b) but with ionizing photon rates
measured by scaling a standard quasar \sed\ to the 2--10~keV
luminosity of the AGN.
\label{F:QionQHa}}
\end{figure*}


A complementary test is depicted in Figure~\ref{F:QionQHa}a, where we
compare the ionizing photon rates of the weak AGNs in our sample of
LINERs to the corresponding H\a\ photon rates. The ionizing photon
rate in this figure was obtained by direct integration of the \sed.
Assuming that H\a\ photons are produced by case~B recombination, one
H\a\ photon is emitted for every 2.2 recombinations
\citep{osterbrock89}. Collisional excitation of H\a\ is unlikely to
make a significant contribution to the observed H\a\ photon rate since
the densities in the emission-line nebulae of LINERs are thought to be
relatively low.  Therefore, the minimum requirement for photon
balance, if all ionizing photons are absorbed by the emission-line
nebula, is $Q_{\rm 1\;Ry-100\;keV} > 2.2\; Q_{\rm H\alpha}$. This
condition is shown as a solid line in Figure~\ref{F:QionQHa}a. Only
14/35 LINERs satisfy this condition. As we have argued earlier in this
section, it is possible that only a fraction $f_{\rm c}$ of the
ionizing photons are absorbed by the emission-line nebula. Thus a more
appropriate form of the photon balance condition is $Q_{\rm
  1\;Ry-100\;keV} > 2.2\; Q_{\rm H\alpha} / f_{\rm c}$, which is
represented in Figure~\ref{F:QionQHa}a as a dashed line, for $f_{\rm
  c}=0.1$. This stricter requirement is satisfied by only 4/35 LINERs
in our sample.  Figure~\ref{F:QionQHa}b is analogous to
Figure~\ref{F:LionLHa}b; it shows the distribution of the ratio
$Q_{\rm 1\;Ry-100\;keV} / Q_{\rm H\alpha}$ with the solid and dashed
lines having the same meaning as in
Figure~\ref{F:QionQHa}a. Figures~\ref{F:QionQHa}c,d show the same test
as Figures~\ref{F:QionQHa}a,b, but with the ionizing photon rates
determined by scaling the standard quasar \sed. The conclusions from
Figure~\ref{F:QionQHa}c are the same as those from
Figure~\ref{F:QionQHa}a; only 15/35 LINERs satisfy the minimum photon
balance condition and only 5/35 objects satisfy the stricter
condition, assuming $f_{\rm c}=0.1$

\section{Summary and Discussion}\label{S:discussion}

The energy and photon budget tests presented in the previous section
show that the weak AGNs in LINERs are unable to power the luminosity
measured in the 2\asec$\times$4\asec\ spectroscopic slit by
photoionization. The AGN can provide enough ionizing photons to
explain the observed H\a\ luminosity in only 43\% of the LINERs of our
sample, if all the available ionizing photons are absorbed by the
emission-line gas. More realistically, however, if only a fraction of
the available ionizing photons are absorbed by the emission-line
nebula, then only in a small minority of cases is the AGN ionizing
photon rate sufficient to explain the observed H\a\ photon rate. In
approximately 1/3 of the cases the ionizing photon rate from the AGN
falls short of the required rate by an order of magnitude or more.

Our conclusion is in agreement with most of the previous studies (see
\S\ref{S:intro}), even though we have assumed a different shape for
the \sed. The main improvement in our approach is that we have assumed
AGN \seds\ that are not simple power laws. The observed \seds\ that we
used can be often described by a relatively steep power law from
2500~\AA\ to 0.5~keV and a flatter power-law from 0.5 to 10~keV (see,
for example, Figs.~2 and 7 of \pone). Our template \seds\ include a
``big blue bump,'' also resulting in a steep slope in the extreme-UV
band, which then becomes flatter in the X-ray band.  In contrast,
\citep{flohic06} extrapolated the power law from the X-ray band to
lower energies and probably underestimated the ionizing luminosity by
missing a substantial fraction of the UV photons. On the other hand,
\citep{maoz95} used a power law \sed\ normalized to the UV flux, thus
probably overestimating the ionizing luminosity.

\subsection{Have We Missed Any Ionizing Photons from the Little Monster?}\label{S:census}

Before we consider other processes that contribute to the ionizing
photon budget, we ask whether the ionizing luminosities and photon
rates that we have determined for the AGNs in question represent the
true values.

\begin{itemize}

\item
{\it Could the ionizing continuum that we detect be extinguished, thus
  weaker than what illuminates the line-emitting gas?} We consider
this an unlikely possibility since we have made generous corrections
for extinction and have tried to err on the side of caution. One of
the methods we used to obtain the ionizing luminosity, scaling the
standard quasar \sed\ to the observed 2--10~keV luminosity, bypasses
this problem. It relies only on a measurement of the X-ray flux which
can be reliably corrected for extinction since the column is
determined directly from the shape of the X-ray spectrum.  Two
additional pieces of evidence bolster this view: (a) the morphology of
the emission-line regions of LINERs is rarely bipolar, disfavoring the
presence of obscuring tori of the type invoked in Seyfert galaxy
unification schemes \citep{pogge00}, and (b) if the ionizing continuum
were considerably stronger than what we observe, the emission line
ratios would have been different, i.e., they would have resembled
those of Seyferts galaxies, indicating of a higher ionization level
of the line-emitting gas.

\item
{\it Could there be an unseen far-UV ``bump'' in the \sed\ which makes
up the power deficit?} This effect is also captured by our method of
obtaining the X-ray luminosity by scaling the standard quasar \sed,
since the template already includes a far-UV bump. If the far-UV bump
in the \sed\ of ``little monsters'' were stronger than that of
quasars, it would have revealed itself in the form of a larger
absolute value of \aox\ and/or a soft X-ray excess, above the
high-energy power-law shape; neither of these features is observed.
Moreover, a stronger far-UV bump would have also produced
Seyfert-like rather than LINER-like emission-line ratios.

\item
{\it Could we be observing the ``echo'' of a previous epoch, a few
  hundred years ago, when the ``little monster'' was more ferocious?}
This possibility was considered by \citet*{eracleous95}, who showed
that the reverberation of an ionizing flare in the nebula produces
LINER-like emission-line ratios. However, this cannot be a universal
explanation. This mechanism requires the flare recurrence times to be
of order a few hundred years and the duty cycle to be relatively short
(i.e., of order a few tens of percent). However, the duty cycle cannot
be arbitrarily short because that would affect the diagnostic line
ratios that are used to classify an object as a LINER. More
specifically, the \flion{O}{3}{5007}$\;/\;$H\b\ ratio, which is often
used as a diagnostic, is very sensitive to the duty cycle because of
the very short decay time of the \flion{O}{3}{5007} line following the
decay of the ionizing continuum. The required combination of
recurrence times and duty cycles of such flares suggests that at least
some of the central UV sources of LINERs should be observed to decay
systematically over the course of a few decades. But this is in
contradiction with the observations of \citet{maoz05} and
\citet{maoz95}, which show the UV sources to persist over time scales
on the order order of a decade. Therefore we are led to disfavor this
particular scenario based on this particular combination of duty cycle
and recurrence time. We also note an additional observational test
that bolsters our conclusion. The echo of an ionizing continuum flare
should be detectable in a narrow-band \flion{O}{3}{5007} image in the
form of a ring-like structure. But \cite{pogge00} have specifically
looked for such structures in their emission-line imaging survey and
did not find convincing evidence for them.

\end{itemize}

\begin{figure}
\centerline{\includegraphics[scale=0.45,angle=0]{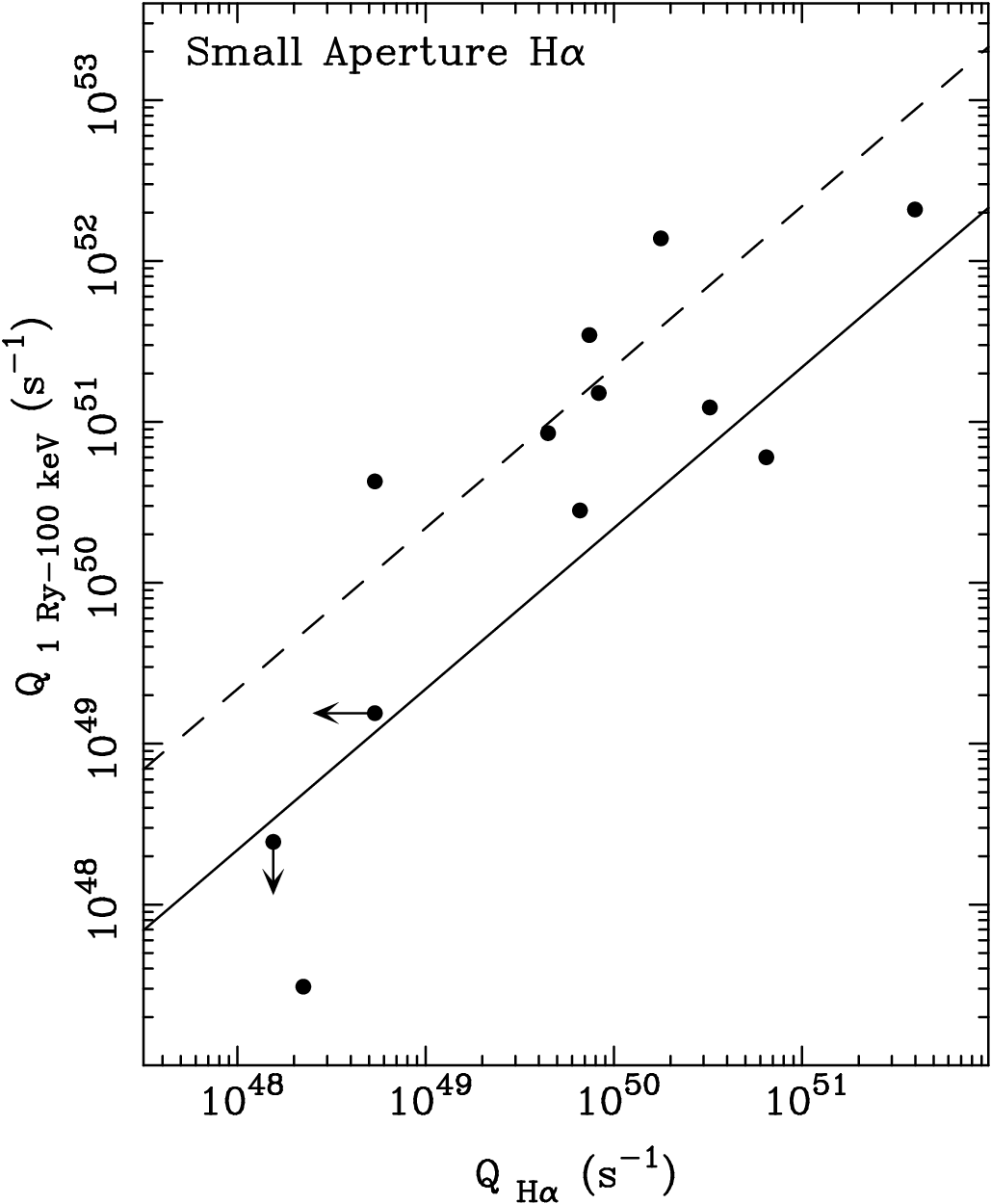}}
\caption{The ionizing (1~Ry--100~keV) photon rate of 12 of the LINERs
  in our sample, plotted against their H\a\ photon measured in a small
  aperture. The ionizing photon rate was determined scaling the
  ionizing photon rate of the standard radio-quiet quasar \sed,
  according to the 2--10~keV luminosity of the AGN. The H\a\ photon
  rates were either measured directly from an \hst\ spectrum or
  determined from the ground-based spectrum by scaling according to
  the fluxes measured in narrow-band images (see details in
  Table~\ref{T:smallapp} and in \S\ref{S:discussion} of the
  text). Arrows denote upper limits on the ionizing photon rate or the
  H\a\ photon rate.  The ionizing photon rate balances the H$\alpha$
  photon rate along the solid line ($Q_{\rm 1\;Ry-100\;keV}=2.2\;Q_{\rm
    H\alpha}$; see \S\ref{S:comparison} of the text). If only a
  fraction $f_{\rm c}$(=0.1) of the ionizing photons are absorbed by
  the nebula, then photon number balance is represented by the dashed
  line ($Q_{\rm 1\;Ry-100\;keV}=2.2\;Q_{\rm H\alpha}/f_{\rm c}$).
\label{F:QionQHasmall}}
\end{figure}

The ``little monster'' is still capable of powering the line emission
from a compact region immediately surrounding it. In fact, the
narrow-band images of \citep{pogge00} show centrally concentrated
emission regions. To illustrate this point, we have collected in
Table~\ref{T:smallapp} measurements of the H\a\ luminosity of 12
objects made with the \hst\ through apertures of angular sizes of a
few tenths of arc-seconds (after making extinction corrections). These
measurements come from STIS spectra \citep{shields07} and WFPC2
narrow-band images \citep{pogge00}. For another two objects, we
measured the line fluxes ourselves from archival STIS and FOS
spectra. In the case of the narrow-band images, the filter includes
both the H\a\ line as well as the [\ion{N}{2}] doublet that straddles
it. We have rescaled the flux according to the relative intensities of
these lines reported by \citet{ho97b}. Adopting the relative
intensities from the large-aperture spectrum is justified by the
results of \cite{shields07}, who find that in the majority of objects
in their sample, the value of the [\ion{N}{2}]/H\a\ ratio measured in
the narrow STIS aperture is within a factor of 1.6 of the value
measured in the large-aperture spectrum of \citet{ho97b}.  In
Figure~\ref{F:QionQHasmall} we plot the ionizing photon rate [from
  column (7) of Table~\ref{T:master}] against the small-aperture
H\a\ photon rate for these 12 objects in order to assess their energy
budgets following the methodology of \S\ref{S:comparison}.  Examining this
figure we see that in 9 out of the 12 objects the ionizing photon rate
is at least several times higher than the rate required to power the
H\a\ lines. Thus, the compact emission-line regions measured through
the small \hst\ apertures could well be powered by the ``little
monster.'' However, this test also underscores our earlier conclusion
that the ``little monster'' can power only a small fraction of the
line luminosity measured through the large,
2\asec$\times$4\asec\ aperture.

\begin{deluxetable}{lrrlc}
\tabletypesize{\scriptsize}
\tablewidth{3.3in}
\tablecaption{H$\alpha$ Luminosities Through Small Apertures\label{T:smallapp}}
\tablehead{
\colhead{Galaxy} &
\colhead{$\log L_{\rm H\alpha}\;$\tablenotemark{a}} &
\colhead{$Q_{\rm i}/Q_{\rm H\alpha}\;$} &
\colhead{Aperture Size} &
\colhead{Notes} \\
\colhead{(1)} &
\colhead{(2)} &
\colhead{(3)} &
\colhead{(4)} &
\colhead{(5)} 
}
\startdata
NGC 0404 & $ 36.83$ & $ 0.14$ & $R=0\farcs23$             & \tablenotemark{b}   \\
NGC 1097 & $ 38.35$ & $46.16$ & $0\farcs25\times0\farcs2$ & \tablenotemark{c}   \\
NGC 3031 & $ 38.40$ & $18.07$ & $0\farcs25\times0\farcs2$ & \tablenotemark{c}   \\
NGC 3998 & $ 40.08$ & $ 5.15$ & $R=0\farcs23$             & \tablenotemark{b}   \\
NGC 4143 & $ 38.13$ & $19.02$ & $0\farcs25\times0\farcs2$ & \tablenotemark{d}   \\
NGC 4314 & $ 36.67$ & $<1.58$ & $0\farcs25\times0\farcs2$ & \tablenotemark{d}   \\
NGC 4374 & $ 38.30$ & $ 4.24$ & $R=0\farcs23$             & \tablenotemark{b}   \\
NGC 4486 & $ 38.99$ & $ 3.85$ & $R=0\farcs23$             & \tablenotemark{b}   \\
NGC 4548 & $ 37.21$ & $79.75$ & $0\farcs25\times0\farcs2$ & \tablenotemark{d}   \\
NGC 4579 & $ 38.73$ & $77.60$ & $R=0\farcs23$             & \tablenotemark{b}   \\
NGC 4594 & $ 39.29$ & $ 0.93$ & $R=0\farcs23$             & \tablenotemark{b}   \\
NGC 5055 & $<37.21$ & $>2.89$ & $0\farcs25\times0\farcs2$ & \tablenotemark{d}   \\
\enddata
\tablenotetext{a}{The luminosities are measured in \ergs. They have
  been corrected for extinction using the values of $E(B-V)$ form the
  emission-line gas reported by \citet{ho97b} and the \citet{seaton79}
  extinction law.}
\tablenotetext{b}{Taken from the narrow-band imaging survey of \cite{pogge00}.
                  Since, the filter includes the H\a\ line and
                  [\ion{N}{2}] doublet, the flux was rescaled
                  according to the relative strengths of the lines
                  measured in the ground-based spectra
                  of \citet{ho97b}.}
\tablenotetext{c}{Measured from archival \hst\ spectrum}
\tablenotetext{d}{Taken from \citet{shields07}}
\end{deluxetable}

The ionizing-photon deficit that we have identified in LINERs should
not extend to more luminous AGNs, even though this deficit appears
even when we assume a quasar-like \sed. More luminous AGNs should emit
more ionizing photons per unit H\a\ luminosity than LINERs based on
the following two indicators. (a) A close examination of Figure~7 of
\citet{flohic06} shows that most LINERs fall below the extrapolation
of the $L_{\rm 2-10\;keV}$-$L_{\rm H\alpha}$ correlation defined by
more luminous AGNs \citep[shown in Fig.\ 2 of][]{ho01}. This means
that the $Q_{\rm i}/Q_{\rm H\alpha}$ ratio for more luminous AGNs will
be higher than that of the weak AGNs in LINERs. (b) For luminous AGNs
with $(\nu L_\nu)_{\rm 2500\;\AA}\gs 10^{43}~\ergs$ the values of
\aox\ are higher than those of the weak AGNs in LINERs \citep[see
  Fig.\ 2 of][]{maoz07}. This implies that luminous AGNs emit more
ionizing photons per unit X-ray luminosity than the weak AGNs in
LINERs.

\subsection{Alternative Sources of Power}\label{S:altpower}

Thus we are led to seek other power sources that could make up
the power deficit in LINERs. These power sources fall in two broad
categories, sources of ionizing radiation that are not associated with
the ``little monster'' and sources of mechanical heating, which may or
may not be associated with the ``little monster.''  Suitable power
sources must fulfill two conditions: (a) they must be able to provide
the requisite power (at least in principle), and (b) they must lead to
LINER-like emission-line ratios.

\begin{itemize}

\item
{\it Mechanical Power Delivered by Compact Jets. --} A number of
authors have shown that the ``little monsters'' found in LINERs are
associated with radio sources with a high brightness temperature
and/or elongated radio morphology \citep[e.g.,][]{nagar01, ulvestad01,
anderson04, filho04, nagar05}. This has been interpreted as evidence
for jets, which give rise to the bulk of the observed radio
emission. \citet{nagar05} have estimated that the kinetic luminosity
of the jets can exceed the 2--10~keV X-ray luminosity by as much as
4.5 orders of magnitude. More specifically, in half of the weak AGNs
in the sample of \citet{nagar05} the kinetic power of the jet exceeds
the 2--10~keV X-ray luminosity by at least a factor of 30.  In
comparison, the ionizing luminosity of the ``little monsters'' in our
sample is 5--10 times larger than their 2--10~keV
luminosity \footnote{The standard radio-quiet quasar \sed\ gives
$L_{\rm 1\;Ry-100\;keV}/L_{\rm 2-10\;keV}=10$. Integrating the observed
\seds\ of the 10 objects in our sample which are detected both in the
X-rays and the UV and the UV light is {\it not} dominated by hot stars
(see \pone), we obtain a mean value of this ratio of 5, a geometric
mean of 6 and a median of 9.}. Thus the kinetic power of the jets can
be up to 3 orders of magnitude higher than the ionizing luminosity of
the AGN.  Even if only a small fraction of this power can be used to
energize the emission-line nebula through shocks, it would be enough
to balance the energy budget.

Shock excitation models have had mixed success in explaining the
emission-line spectra of LINERs. On the positive side,
\citet{nicholson98} find that the optical+UV spectrum from the nucleus
of NGC~3998 can be explained by models invoking slow shocks
\citep[$\ls 150$~\kms; see][]{shull79}. On the negative side, however,
\citet*{ho96} have shown that the optical+UV spectrum of the nucleus
of M81 is not compatible with shock models. More generally,
\citet{maoz98} pointed out that none of the UV spectra of the seven
LINERs in their sample resembles those predicted by shock models (they
either have broad, AGN-like emission lines or they resemble the
spectra of starbursts, lacking strong emission lines). This is also
the case for the nuclear UV spectrum of NGC~1097 \citep{storchi05}.

Perhaps the role of shocks can be better understood by considering
the case studies of the nucleus and circumnuclear disk of M87 based on
spectra obtained through small (sub-arcsecond) apertures with the
\hst.  According to \citet{dopita96b} and \citet{dopita97} the
optical+UV emission-line spectrum of the circumnuclear gaseous disk of
M87 can be explained by models invoking fast ($\gs 200$~\kms) shocks.
After studying the inner regions of the disk of M87, \citet{sabra03}
conclude that the gas is excited by photons from the active
nucleus but they also note that there appears to be a transition from
photoionization to shock excitation at a distance of a few tens of
parsec from the center.

\item
{\it Photoionization by Young Stars. --} Young star clusters are found
in some LINERs, although they are not very common. For example,
\citet{maoz98} find the signature of hot stars in the UV spectra of
3/7 LINERs in their sample (NGC~404, NGC~5055, and NGC~4569), while by
similar means, \citet{storchi05} find a compact ($< 9$~pc)
star-forming region around the AGN in NGC~1097. We can make rough
estimates of the incidence of young stellar populations in the nuclei
of LINERs by consulting the results of \citet{cid05} and
\citet{gonzalez04} who have used spectral decomposition techniques to
study the optical and UV spectra of LINERs and ``transition objects''.
Based on these considerations, we find that $\sim 10$--15\% of LINERs
harbor young stellar populations in their nuclei.

When present, young star clusters may be able to provide enough power
to balance the energy budget. In fact, \citet{maoz98} have
demonstrated that this is possible for the 3 LINERs with
starburst-like UV spectra out of their sample of 7.  They used
synthetic spectra of young star clusters normalized to the observed UV
luminosities to show that there are enough ionizing photons in the
unobservable far-UV band to account for the observed emission-line
luminosities (assuming that all ionizing photons are absorbed by the
emission-line nebula). We also note that photoionization by extremely
hot, massive stars can explain the relative intensities of the optical
emission lines, as shown by the models of \citet{terlevich85},
\citet{filippenko92}, \citet{shields92} and \citet{barth00}. In
addition to photoionization, shocks from supernovae or stellar winds
in a young cluster could also contribute to the excitation of the
emission-line gas.

\item
{\it Photoionization by Post-AGB Stars from an Old or Intermediate-Age
  Stellar Population. --} This idea was suggested by \citet{binette94}
  to explain the line emission from normal elliptical galaxies, but
  the authors also noted that the same scenario may be applicable to
  LINERs.  A variant of this idea, in which the ionizing photons are
  provided by the central stars of planetary nebulae was explored by
  \citet{taniguchi00}.  Using a population synthesis simulation,
  \citet{binette94} computed the evolution of the ionizing photon rate
  from an aging stellar population and found that at late times (after
  the demise of the hot, massive stars at $t\gs 10^8$ years) the
  dominant contribution to the ionizing luminosity comes from post-AGB
  stars. They also found that the ionizing photon rate at late times
  declines fairly slowly with time (it drops by an order of magnitude
  or less between the ages of $10^8$ and $10^{10}$ years). By carrying
  out a series of photoionization calculations, they demonstrated that
  the resulting emission line spectra resemble those observed in
  LINERs. This scenario is attractive because it relies only on
  ionizing photons from the old stellar population. It is also
  bolstered by one of the results of the X-ray survey of LINERs by
  \cite{flohic06}: the H\a\ luminosity is more tightly correlated to
  the soft (0.5--2~keV) than to the hard (2--10~keV) X-ray luminosity
  of the nucleus, which is likely to be associated with stellar
  processes (e.g., stellar winds).

\begin{deluxetable*}{cccccccc}
\tabletypesize{\scriptsize}
\tablewidth{0in}
\tablecaption{Ionizing Photons from Post-AGB Stars\label{T:postAGB}}
\tablehead{
\colhead{} &
\colhead{$B-V$\tablenotemark{a}} &
\colhead{$B_{\rm slit}$\tablenotemark{b}} &
\colhead{} &
\colhead{} &
\colhead{} &
\colhead{} &
\colhead{} \\
\colhead{Galaxy} &
\colhead{(mag)} &
\colhead{(mag)} &
\colhead{$\log (M_{\star}/{\rm M}_{\odot})$\tablenotemark{c}} &
\colhead{$\log Q_{\rm \star,\mu_B}$\tablenotemark{d}} &
\colhead{$Q_{\rm \star,\mu_B}/Q_{\rm H\alpha}$\tablenotemark{e}} &
\colhead{$Q_{\rm \star,\mu_B}/Q_{\rm AGN}$\tablenotemark{f}} &
\colhead{Refs.\tablenotemark{g}} \\
\colhead{(1)} &
\colhead{(2)} &
\colhead{(3)} &
\colhead{(4)} &
\colhead{(5)} &
\colhead{(6)} &
\colhead{(7)} &
\colhead{(8)} 
}
\startdata
NGC 3379  &  0.98  &  13.81  &  9.64  &  50.50  &  11.02  &   234    & 1 \\
NGC 3607  &  0.96  &  14.50  &  9.37  &  50.24  &   0.61  &  $>1.7$  & 1 \\
NGC 3608  &  0.98  &  14.82  &  9.24  &  50.10  &   2.00  &  $>2.7$  & 1 \\
NGC 4278  &  0.97  &  14.63  &  9.31  &  50.18  &   0.31  &    0.2   & 1 \\
NGC 4374  &  0.99  &  14.96  &  9.35  &  50.22  &   0.65  &    0.6   & 2 \\
NGC 4494  &  0.91  &  14.73  &  9.30  &  50.16  &  12.64  &    2.1   & 1 \\
NGC 4552  &  0.98  &  14.29  &  9.45  &  50.31  &   1.87  &    1.0   & 1 \\
NGC 4636  &  0.95  &  15.16  &  9.11  &  49.97  &   1.53  &    1.9   & 3 \\
\enddata
\tablenotetext{a}{The $B-V$ color of the galaxy, taken from the RC3 
catalog \citep{devaucouleurs91}.}
\tablenotetext{b}{The apparent $B$ magnitude of the area of the galaxy
                  within the 2\asec$\times$4\asec spectroscopic
                  aperture. For NGC~4374, we integrated the $g$-band
                  surface brightness profile and converted to a $B$
                  magnitude using the transformation
                  $B=g+0.47(B-V)+0.07$ \citep{smith02}. For all other
                  galaxies we integrated the $V$-band surface
                  brightness profiles and converted to a $B$ magnitude
                  using the colors reported in this table.}
\tablenotetext{c}{The mass of stars within the area of the slit,
                  estimated from the $B$-band mass-to-light ratio
                  derived by \citet{binette94}, $M/L_{\rm
                  B}=8(M/L_{\rm B})_{\odot}$. We adopted the distances
                  given by \citet{ho97b} so as to compare directly
                  with the H\a\ luminosity.}
\tablenotetext{d}{The ionizing photon rate of the post-AGB stars from
                  the old stellar population ($t\sim 10^{10}$~years) in s$^{-1}$,
                  estimated from the prescription of \citet{binette94}. See
                  details and discussion in \S\ref{S:discussion} of the text.}
\tablenotetext{e}{The ratio of the ionizing photon rate from the
                  post-AGB stars to the H\a\ photon rate. If all
                  ionizing photons are absorbed by the line-emitting
                  gas, a minimum ratio of 2.2 is needed if the H\a\
                  photons are to be produced by case B
                  recombination. }
\tablenotetext{f}{The ratio of the ionizing photon rate from the old
                  stellar population to the ionizing photon rate from
                  the ``little monster.''}
\tablenotetext{g}{References to surface brightness measurements: (1)
  \citet{lauer05}, $V$-band light profiles from \hst/WFPC2 F555W
  images; (2) \citet{ferrarese06}, $g$-band light profiles from
  \hst/ACS F475W images; (3) \citet{lauer95} $V$-band light profiles
  from \hst/WFPC F555W images.}
\end{deluxetable*}

To investigate the plausibility of this scenario, we have used the
surface brightness profiles of eight elliptical or S0 galaxies from
our sample available in the literature. These objects, listed in
Table~\ref{T:postAGB}, were observed in bands resembling the Johnson
$V$ or $B$ bands with the \hst's WFPC, WFPC2, or ACS. Thus, we are
able to estimate whether there are enough post-AGB stars within the
2\asec$\times$4\asec\ aperture of the spectroscopic observations
\citep*{ho95} to provide the requisite number of ionizing
photons. After integrating the surface brightness over the area of the
spectroscopic aperture and applying the appropriate color corrections
(see details in Table~\ref{T:postAGB} and associated footnotes) we
derived the slit $B$ magnitudes. The population synthesis model of
\citet{binette94} gives a mass-to-light ratio of $M/L_{\rm B} =
8(M/L_{\rm B})_{\odot}$ and a specific ionizing photon rate from
post-AGB stars of $dQ_{\star}/dM = 7.3\times 10^{40}~{\rm s^{-1}
  ~M_{\odot}^{-1}}$, allowing us to estimate the total ionizing photon
rates from post-AGB stars within the spectroscopic aperture, $Q_{\rm
  \star,\mu_B}$ (where the subscript $\mu_{\rm B}$ indicates that
$Q_\star$ was obtained from the galaxy's surface brightness
profile). We summarize our results in Table~\ref{T:postAGB}, where we
also compare $Q_{\rm \star,\mu_B}$ to the H\a\ photon rate, $Q_{\rm
  H\alpha}$, and the ionizing photon rate from the ``little monster,''
$Q_{\rm AGN}$.

An alternative way of obtaining $Q_\star$ \citep[used by][]{ho08}
employs the spectroscopic B magnitude, $m_{44}$, measured and reported
by \citet{ho97b}. After transforming $m_{44}$ to a standard B
magnitude via $m_{\rm B}=m_{44}+0.140\;$\footnote{The slit magnitude
  is defined as $m_{44}=-2.5\log f_\nu-48.6$ and it is determined from
  the spectra by averaging the flux density in two narrow windows
  straddling 4300~\AA: 4262--4281~\AA\ and
  4360--4375~\AA\ \citep{ho97b}. In comparison, the standard B
  magnitude has a bandpass with an effective wavelength of
  4300~\AA\ and it is defined such that $m_{\rm
    B}=-2.5\log(f_\nu/4130\;{\rm Jy})$.}, we apply the procedure of
the previous paragraph. We denote the ionizing photon rate obtained in
this manner as $Q_{\star,m_{44}}$ and we include the resulting values
in Table~\ref{T:master} for the galaxies for which $m_{44}$ is
available. The method involving $m_{44}$ is subject to the uncertainty
that the B magnitude is not measured in a standard manner. Therefore,
we have compared the the resulting values of $Q_{\star,m_{44}}$ and
$Q_{\rm \star,\mu_B}$ for all the objects of Table~\ref{T:postAGB} and
we have found that the ratio $Q_{\star,m_{44}}/Q_{\rm \star,\mu_B}$
has a mean value of 1 and a standard deviation of 0.6, indicating that
the agreement between the two methods is reasonably good.

An examination of Table~\ref{T:postAGB} shows that in 2/8 cases the old
stellar population provides enough ionizing photons to power the
observed H\a\ emission, even if $\sim 20$\% of these photons are
absorbed by the gas. In 3/8 cases the H\a\ emission could be powered
by ionization from the old stellar population only if all the ionizing
photons were absorbed by the gas, while in the remaining 3/8 cases,
the rate of ionizing photons provided by the old stellar population falls
short of the requisite rate by a factor of a few. A consideration of
the values of $Q_{\star,m_{44}}$ listed in Table~\ref{T:master} lead
to similar conclusions. In Figure~\ref{F:QstarQHa} we show a histogram
of the ratio $Q_{\star,m_{44}}/Q_{\rm H\alpha}$ to illustrate this
result. In 4/28 objects, the post-AGB stars provide enough ionizing
photons to account for the observed H\a\ luminosity, in 5/28 objects
the H\a\ luminosity may be powered if all the ionizing photons from
the post-AGB stars are absorbed by the emission-line gas, and in the
remaining 19/28 objects, the ionizing photons from the post-AGB stars 
cannot produce the observed H\a\ luminosity.

\begin{figure}
\centerline{\includegraphics[scale=0.45,angle=0]{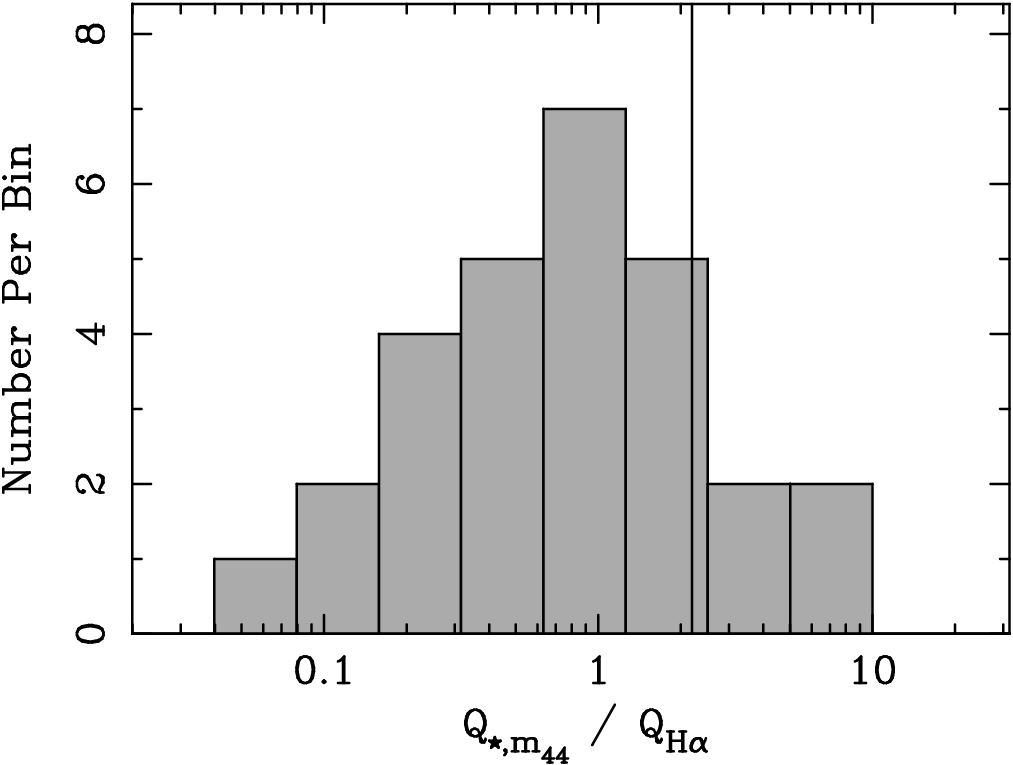}}
\caption{\label{F:QstarQHa} Distribution of the ratio of rates of
  ionizing photons from post-AGB stars to H\a\ photons in the
  2\asec$\times$4\asec\ spectroscopic aperture. The rate of ionizing
  photons was computed from $m_{44}$, the spectroscopic B magnitudes
  reported by \citep{ho97b}, using the methodology described in \S5 of
  the text. The vertical line indicates $Q_{\star,m_{44}}/Q_{\rm
    H\alpha}=2.2$, the minimum requirement for photoionization by
  post-AGB stars to power the observed H\a\ luminosity.}
\end{figure}

It is possible that the available ionizing photon rate is higher than
what we have estimated if an intermediate-age population is present.
The spectroscopic decomposition analysis of \citet{cid05} and
\citet{gonzalez04} suggest that in $\sim 30$\% of LINERs and
``transition objects'' a significant fraction of the optical and/or UV
light comes from a population of intermediate age ($t\sim
10^8-10^9$~years). In such a case, the ionizing photon rates from
post-AGB stars can be up to an order of magnitude higher than the
values listed in Tables~\ref{T:master} and \ref{T:postAGB} and they
would be enough to power the observed H\a\ luminosities in a large
fraction of LINERs.

We also note that in the majority of galaxies of Table~\ref{T:postAGB}
the ionizing photon rates from the old stellar population exceed the
ionizing photon rates from the respective AGNs. Similarly, in 14/29
galaxies in Table~\ref{T:master}, $Q_{\star,m_{44}}/Q_{\rm i} > 1$
(using the value of $Q_{\rm i}$ determined by scaling the radio-quiet
quasar \sed). This suggests that the old or intermediate-age stellar
population can at least be as important a source of power for the
emission-lines as the AGN itself.

\end{itemize}

\subsection{Conclusions}\label{S:conclusions}

The above considerations lead us to conclude that mechanical heating
from the AGN and/or photoionization from the {\it old} stellar
population are important power sources in a large fraction of LINERs
on scales of order 200~pc from the nucleus (corresponding to the
2\asec$\times$4\asec\ aperture at the typical distances of our
targets), although they may not be universal. Photoionization from the
AGN can power the emission from gas on scales of order a few tens of
parsec but it seems to be an inadequate source of power for gas on
larger scales in the majority of cases. Young stellar populations seem
to be rare but important when present. Our conclusions regarding the
importance of post-AGB stars are in agreement with the conclusions of
\citet{stasinska08} who study the stellar populations and excitation
mechanisms of the gas in a large sample LINERs selected from the Sloan
Digital Sky Survey. The emission lines fluxes for that study were
measured through an apperture of of the same area as the fluxes that
we have used here. Furthermore, using data from the SAURON project,
\citet{sarzi10} have pusued the question of the power source of the
nebular emission in the nuclei of elliptical and lenticular galaxies
(on angular scales from a few to 15 arc-seconds) and have concluded
that post-AGB stars most promising possibility.  On the negative side,
however, as noted by \citet{sarzi10}, it is not clear whether
population synthesis models provide reliable estimates of the numbers
of post-AGB stars in an old stellar population. This is undersored by
the recent UV imaging survey of M32 by \citet{brown08}, who find
considerably fewer post-AGB stars than expected.  In conclusion, it
appears likely that we are observing hybrid emission-line spectra
powered by a combination of physical processes, with different
processes potentially dominating on different scales.  A similar
conclusion was reached by \citet{sabra03} in their case study of M87.

\begin{figure}
\centerline{\includegraphics[scale=0.45,angle=0]{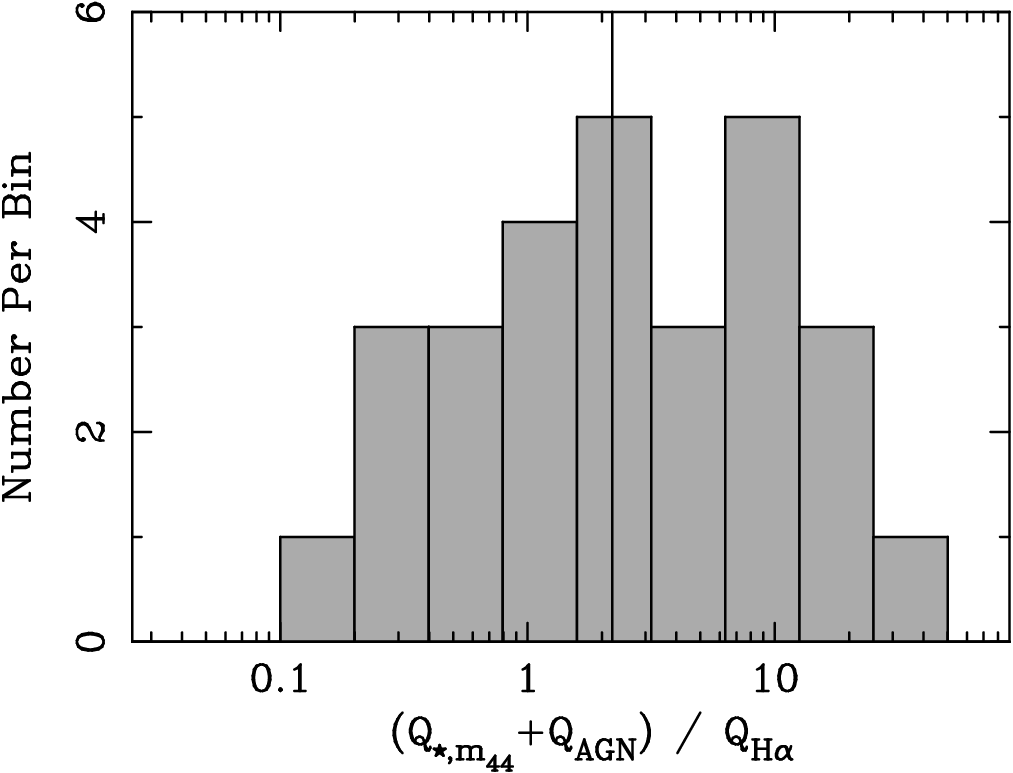}}
\caption{\label{F:QstarAGNQHa} Distribution of the ratio of rates of
  ionizing photons from the AGN {\it plus} post-AGB stars to H\a\
  photons in the 2\asec$\times$4\asec\ spectroscopic aperture. The rate
  of ionizing photons from post-AGB stars was computed from $m_{44}$,
  the spectroscopic B magnitudes reported by \citep{ho97b}, using the
  methodology described in \S5 of the text. The rate of ionizing
  photons from the AGN was computed by scaling the average radio-quiet
  quasar SED. The vertical line indicates $(Q_{\star,m_{44}}+Q_{\rm
    AGN})/Q_{\rm H\alpha}=2.2$, the minimum requirement for
    photoionization to power the observed H\a\ luminosity.}
\end{figure}

Nevertheless, we have yet to balance the energy budget of a significant
fraction of LINERs, thus we cannot uniquely identify the mechanisms
that give rise to the classic LINER spectrum on scales of order a few
hundred parsec around the nucleus. To illustrate this point, we compare
the number of ionizing photons contributed by the combination of the
AGN and the old stellar population with the number of emitted
H\a\ photons. In Figure~\ref{F:QstarAGNQHa} we plot the distribution
of the ratio $(Q_{\star,m_{44}}+Q_{\rm AGN})/Q_{\rm H\alpha}$ and we
mark the value of 2.2, which is the minimum requirement for a balanced
photon budget. As Figure~\ref{F:QstarAGNQHa} shows, there is a deficit
of ionizing in approximately half of the targets, even if we add the
contributions from the AGN and the post-AGB stars. This suggests that
there is at least one important power source that remains
unidentified.

Perhaps the most direct way to address the question of the power
source of LINERs is optical+UV spectroscopy at high spatial
resolution. The 1200--3200~\AA\ UV band is desirable because it
includes important diagnostic emission lines that can distinguish
between shocks and photoionization. It is also the best band to search
for the signature of massive stars. High spatial resolution is
desirable in order to map out the nuclear region spectroscopically and
investigate whether different mechanisms are responsible for exciting
the emission-line gas at different distances from the center. With
such UV spectra it may be possible to assess the role of shocks in
powering the emission line regions on scales of order a few hundred
parsec by comparing the relative intensities of weak optical and UV
lines with th predictions of shock models \citep[see, for
  example][]{dopita96b,wilson99}. Moreover, the photoionizing shock
models of \citet{dopita96b} imply that there should be a diffuse
source of ionizing continuum embedded in the line-emitting gas on
scales of order a few hundred parsec, which may be detectable.

\acknowledgements 

We thank the referee, D. Maoz for many helpful comments and E. C. Moran
for a critical reading of the manuscript.  We also thank R. A. Wade
and R. B. Ciardullo for very illuminating discussions.  This work was
partially supported by the National Aeronautics and Space
Administration through \chandra\ award number AR4-5010A issued by the
\chandra\ X-Ray Observatory Center, which is operated by the
Smithsonian Astrophysical Observatory for and on behalf of the
National Aeronautics and Space Administration under contract
NAS8-03060. ME acknowledges partial support from the Theoretical
Astrophysics Visitors' Fund at Northwestern University and thanks the
members of the group for their warm hospitality during his stay.  This
research has made extensive use of the NASA/IPAC Extragalactic
Database (NED) which is operated by the Jet Propulsion Laboratory,
California Institute of Technology, under contract with the National
Aeronautics and Space Administration.


\end{document}